\documentclass[twocolumn]{autart}    

\makeatletter
\let\theoremstyle\@undefined  
\makeatother

\usepackage{graphicx}          
\usepackage{lineno}
\usepackage[hypertexnames=false]{hyperref}
\usepackage{amsmath, amsthm, amssymb, amsfonts}
\usepackage{tikz}
\usepackage{pgfplots}
\usepackage{caption}
\usepackage{subcaption}
\usepackage{xcolor}
\usepackage{balance}
\usepackage{algorithm}
\usepackage{algpseudocode}
\usepackage{mathtools}
\usepackage{enumitem}
\usepackage{comment}
\usepackage{tabularx}
\usepackage{bbm}
\usepackage{cite}
\usepackage{caption}
\usepackage{subcaption}
\usepackage{soul}

\newif\ifshowappendixcite

\showappendixcitefalse

\newcommand{\togglecomment}[1]{\ifshowappendixcite#1\fi}

\allowdisplaybreaks

\definecolor{darkgreen}{RGB}{0,100,0}

\DeclareMathOperator*{\argmin}{arg\,min}

\begin{document}

\def\Nick#1{{\textcolor{blue}{ {\bf Nick:} #1}}}

\begin{frontmatter}
\runtitle{Conformal Predictive Programming}  
\newtheorem{theorem}{Theorem}[section]
\newtheorem{proposition}[theorem]{Proposition}
\newtheorem{lemma}[theorem]{Lemma}
\newtheorem{corollary}[theorem]{Corollary}

\newtheorem{definition}[theorem]{Definition}
\newtheorem{assumption}[theorem]{Assumption}
\newtheorem{remark}{Remark} 

\title{Conformal Predictive Programming for Chance Constrained Optimization}

\thanks[footnoteinfo]{These authors contributed equally to this work. This work was supported in part by Viterbi School of Engineering Fellowship, Annenberg Fellowship, and the NSF award SLES-2417075.
M.S.~was partly supported by the NSF (award DMS 2210637) and the USC-Capital One Center for Responsible AI Decision Making in Finance. }

\author[1]{Yiqi Zhao$^\star$}\ead{yiqizhao@usc.edu},    
\author[1]{Xinyi Yu$^\star$},               
\author[1,2]{Matteo Sesia},
\author[1]{Jyotirmoy V. Deshmukh},  
\author[1,3]{Lars Lindemann}

\address[1]{Thomas Lord Department of Computer Science, University of Southern California}
\address[2]{Department of Data Sciences and Operations, University of Southern California}
\address[3]{Automatic Control Laboratory, ETH Z\"{u}rich}
          

\begin{abstract}                          
We propose \textit{conformal predictive programming} (CPP), a framework to solve chance constrained optimization problems, i.e., optimization problems with constraints that are  functions of random variables. CPP utilizes samples from these random variables along with the quantile lemma - central to conformal prediction - to transform the chance constrained optimization problem into a deterministic problem with a quantile reformulation. CPP's main strength is an independent calibration step that provides \textit{a posteriori}  guarantees for the solution of this problem that are of \textit{conditional} and \textit{marginal} nature otherwise. These guarantees even apply in settings when assumptions required for obtaining standard a priori guarantees (e.g., in scenario optimization or sample average approximation) are unavailable, difficult to compute, or conservative. Another strength of CPP is that it can easily support different variants of conformal prediction which have been (or will be) proposed within the conformal prediction community. To illustrate this, we present \textit{robust CPP} to deal with distribution shifts in the random variables and \textit{Mondrian CPP} to deal with class conditional chance constraints. In a series of case studies, we show the validity of the aforementioned approaches, and illustrate the advantage of CPP as compared to scenario approach.
\vspace{-15pt}
\end{abstract}

\end{frontmatter}
\section{Introduction}
We are interested in chance constrained optimization (CCO) problems, which arise in robot navigation \cite{blackmore2011chance}, portfolio optimization \cite{pagnoncelli2009computational}, and control/planning \cite{celik2019chance}. To give a concrete example, in motion planning we are often interested in minimizing the energy consumption of a robot subject to sensor uncertainty and obstacle avoidance constraints. Solutions to this CCO ensure robot safety with high probability even in presence of the sensor uncertainty. We formalize the notion of CCOs next.

Consider a probability space $(\Omega, \mathcal{F}, \mathbb{P})$ with sample space $\Omega$, $\sigma$-algebra $\mathcal{F}$, and probability measure $\mathbb{P}: \mathcal{F} \rightarrow [0, 1]$. Let $Y: \Omega \rightarrow \mathbb{R}^d$ be a random vector defined over $(\Omega, \mathcal{F}, \mathbb{P})$. For simplicity, in this paper, we  denote the distribution of a random variable $Y$ by $P_Y$, i.e., $Y\sim P_Y$.

\textbf{CCO Problems. }For a user-defined (often small) {\em failure probability} $\delta \in (0, 1)$, we define a CCO problem as:
\begin{subequations}\label{eq:cco}
\begin{align}
    \min_{x \in \mathcal{X}} \quad & J(x)\\
    \textrm{s.t.} \quad & \mathbb{P}(f(x, Y) \le 0) \ge 1 - \delta, \label{eq:chance_constraint_}
\end{align}
\end{subequations}
where the \textit{decision variable} $x \in \mathbb{R}^n$ is constrained to be within a pre-defined \textit{deterministic feasible region} $\mathcal{X} \subseteq \mathbb{R}^n$. Here, $f: \mathbb{R}^n \times \mathbb{R}^d \rightarrow \mathbb{R}$ is a Borel measurable function of $x \in \mathbb{R}^n$ and the random vector $Y$, while $J: \mathbb{R}^n \rightarrow \mathbb{R}$ is a \textit{cost function}. We refer to \eqref{eq:chance_constraint_} as an {\em individual chance constraint}, or simply {\em chance constraint}. Joint chance constraints refer to probabilistic constraints  $\mathbb{P}(f_i(x, Y) \le 0, \forall i \in \{1, \hdots, q\}) \geq 1 - \delta$ for $q \in \mathbb{N}$. We focus on individual chance constraints, but our framework can be applied to joint chance constraints following union bound or pointwise maximum encodings \cite{geng2019data}. We also define the \textit{probabilistic feasible region} of \eqref{eq:cco} as $F \coloneq \{x \in \mathcal{X}: \mathbb{P}(f(x, Y) \le 0) \ge 1 - \delta\}$. We denote the \textit{optimal solution} of this problem as $x^* \in \mathcal{X}$. 
As standard in the literature \cite{geng2019data}, we assume that $J(x^*) = -\infty$ if \eqref{eq:cco} is unbounded from below. 
We further assume that $x^*$ exists (i.e., $F \neq \emptyset$) and, without loss of generality, is unique. If $x^*$ is not unique, any tie-breaking rule suffices.

\textbf{Sampling-based approaches. }CCO problems are  difficult to solve because the distribution $P_Y$ is typically unknown in practice. Even in cases where $P_Y$ is known, we have to solve complex (potentially high-dimensional) integrals which is only possible under limiting assumptions on $P_Y$ and the constraint function $f$. To avoid these issues, sampling-based approaches  use samples from $P_Y$ instead, see e.g., \cite{calafiore2006scenario, campi2009scenario, luedtke2008sample}. Such approaches formulate  deterministic optimization problems that use $K$ i.i.d. samples (or scenarios) $Y^{(1)}, \hdots, Y^{(K)}$ from $P_Y$. We review these deterministic optimization problems in the related work section. However, since we are using samples from $P_Y$, we are not always guaranteed to solve~\eqref{eq:cco}. 

For this reason, we are interested in providing guarantees that the solution $x_\text{det}^*$ of the deterministic optimization problem solves the CCO problem  \eqref{eq:cco}. If the solution $x_\text{det}^*$  is a feasible solution to \eqref{eq:cco} with a probability of at least 
$1 - \beta$, we obtain so called \textbf{conditional feasibility guarantees} (or \textbf{PAC guarantees}). These guarantees are conditional as they hold with a confidence of at least $1-\beta$ with respect to $Y^{(1)}, \hdots, Y^{(K)}$. While these guarantees are fairly common, we are also interested in \textbf{marginal feasibility guarantees} in which the solution $x_\text{det}^*$ satisfies $f(x_\text{det}^*, Y) \le C(\mathcal{Y}_\text{cal})$ with a confidence of at least $1-\delta'$ with respect to $Y$ and $\mathcal{Y}_\text{cal}$, where $\mathcal{Y}_\text{cal}$ is a set of samples drawn from $P_Y$. For instance, $\mathcal{Y}_\text{cal}$ could consist of $Y^{(1)}, \hdots, Y^{(K)}$ or a new set of samples $Y^{(K+1)}, \hdots, Y^{(K+L)}\sim P_Y$. Here, one would typically desire that $C\le 0$ and $\delta'=\delta$. Marginal guarantees have so far not been explored in the CCO literature. If the parameters $\beta$ and $\delta'$ are known prior to solving the deterministic optimization problem, one obtains so called \textbf{a priori guarantees}. Obtaining a priori conditional feasibility guarantees for \eqref{eq:cco} typically relies on structural assumptions  of \eqref{eq:cco}  such as convexity \cite{campi2009scenario} or Lipschitz continuity of the constraint functions \cite{luedtke2008sample}. While a priori guarantees are desirable, one can often only determine $\beta$ and $\delta'$  after solving the deterministic optimization problem, which we refer to as \textbf{a posteriori guarantees}. For problems that do not satisfy the aforementioned structural assumptions,  existing approaches such as \cite{campi2018general, garatti2024non, shang2020posteriori} provide a posteriori conditional feasibility guarantees. However, these approaches are either computationally expensive, provide looser feasibility guarantees, or do not generalize to broader classes of CCO problems, e.g., those that are robust to parameter variations. We therefore seek to design a computationally efficient and easy to extend framework for CCO problems with a wide range of statistical guarantees.


\textbf{Contributions.} The main contribution of this paper is the introduction of a new sampling-based approach for solving CCO problems, which we call \emph{conformal predictive programming} (CPP).
CPP leverages conformal prediction (CP), which is a statistical tool for uncertainty quantification that has recently found broad application in autonomous control system and machine learning applications \cite{balasubramanian2014conformal, angelopoulos2021gentle}. Effectively, CPP utilizes samples from $P_Y$ along with the quantile lemma from CP to transform the CCO problem into a deterministic optimization problem. CPP makes limited structural assumptions on the CCO problem and is efficiently solvable. We summarize our contributions as follows:
\vspace{-8pt}
\begin{itemize}[leftmargin=*]
    \item We present CPP as a new framework  for solving CCO problems with limited structural assumptions on \eqref{eq:cco}. We also provide \textbf{conditional} and \textbf{marginal} \textbf{a posteriori} feasibility guarantees and show that CPP inherits \textbf{a priori} feasibility guarantees from existing sample average approximation approaches.
    \item We illustrate the versatility of CPP by incorporating different variants of CP to solve problems beyond CCO, including Robust CCO (RCCO) and our proposed problem of Mondrian CCO (MCCO).
    \item We present case studies to empirically validate CPP. We compare to scenario optimization \cite{campi2018general, garatti2024non}. We further evaluate Robust CPP and Mondrian CPP.
\end{itemize}
\vspace{-8pt}
Due to the space limitation, all proofs and supplementary materials are presented in the Appendix, Section \ref{sec:appendix}. \togglecomment{, which we include in an extended version of the paper \cite{zhao2024conformal}.
\vspace{-8pt}}

\subsection{Related Work}
\label{sec:rel_work}

CCO problems are well studied, with early work dating back to \cite{charnes1959chance}. 
One of the well-known challenges is that, without strong assumptions on $J$, $f$, and $P_Y$, these problems are computationally intractable due to the need to solve complex integrals. 
Early studies addressed this difficulty by assuming specific distributions for the random parameter $Y$ \cite{van1963minimum}. In practice,  these assumptions often do not hold, i.e., $P_Y$ is unknown and non-Gaussian/log-concave \cite{geng2019data}, motivating sampling-based approaches.

\textbf{Scenario Approach (SA).} In SA, we replace the chance constraint \eqref{eq:chance_constraint_} by the deterministic constraint  
$f(x, Y^{(i)}) \le 0$ for  samples $i \in \{1, \hdots, K\}$ to approximate the solution to the CCO problem  \eqref{eq:cco} \cite{calafiore2006scenario, campi2009scenario}. 
If $J$ and $f$ are convex in $x$ and $\mathcal{X}$ is a convex set (referred to as the \textit{convexity assumption} in the remainder), we obtain a priori conditional feasibility guarantees where the confidence $1-\beta$ depends on $K$ and $n$ \cite{campi2008exact, calafiore2010random}.\footnote{Importantly, compared to SA, the guarantees that we present in this paper do not depend on the dimension $n$.}  The sampling-and-discarding variant of SA provides similar guarantees but allows to discard samples from the SA program to increase performance of the solution at the expense of looser feasibility guarantees \cite{campi2011sampling}. Interestingly, one can set up a sampling-and-discarding SA program that recovers conformal prediction guarantees \cite{lin2024verification}. The wait-and-judge variant of SA provides a posteriori conditional guarantees by analyzing the number of support constraints of the solution  \cite{campi2018wait}. This, in many cases, provides tighter guarantees than those from  \cite{campi2008exact}. These approaches are only valid under the convexity assumption and provide guarantees that become more conservative as the number $n$  of decision variables grows, as opposed to our approach. Recent extensions of SA to lift the convexity assumption are presented in \cite{calafiore2012mixed, yang2019chance} under structural assumptions of the CCO. The work in \cite{margellos2014road} lifts the convexity assumption by reformulating the CCO as a robust optimization problem and a convex SA problem. The most general variants of SA are presented in \cite{campi2018general,garatti2024non} where a posteriori conditional feasibility guarantees are obtained. Nonetheless, to obtain tight guarantees, one needs to compute (or find an upper bound of) the smallest set of support samples that maintain the optimal solutions. Evidently, computing this set requires repeatedly solving  optimization problems over subsets of samples. Additionally, in \cite{garatti2024non}, one needs to solve an extra problem with polynomial equality constraints. As in the convex setting, the guarantees in the nonconvex setting depend on the number of decision variables and become looser with large $n$. Our method is practically motivated and, as opposed to SA, uses one dataset for optimization and a new dataset for a posteriori calibration to obtain a practical algorithm that provides guarantees independent of $n$, resulting in less conservatism for large $n$ at the expense of a second dataset.

\textbf{Sample Average Approximation (SAA).} In SAA, the motivation is to use an empirical distribution 
over the set of samples to approximate the chance constraint \eqref{eq:chance_constraint_} directly \cite{luedtke2008sample, luedtke2010integer, pagnoncelli2009sample}. SAA provides a priori feasibility guarantees under specific assumptions such as $\mathcal{X}$ being finite. Additionally, SAA can provide a priori optimality guarantees. We show in Section \ref{sec:cpp} that our CPP approach can be seen as an instantiation of SAA which allows us to inherit the feasibility and optimality guarantees from SAA. Existing quantile based approaches to CCO have developed deterministic reformulations and solution methods for empirical chance constraints \cite{xie2018quantile, pena2020solving, luo2024empirical}. The idea of a posteriori  checking feasibility of an SAA solution was presented in \cite{luedtke2008sample}. However, no finite-sample guarantees were obtained as compared to our work. 

\textbf{Conditional a posteriori feasibility guarantees. } Obtaining a posteriori guarantees has been studied before in the literature. Notably, the authors in  \cite{shang2020posteriori} use the Chernoff bound for convex SA programs. We  establish a connection with our work by showing that our conditional a posteriori feasibility guarantees effectively reduce to those   from a Chernoff bound in \cite{shang2020posteriori}  when the sample size is large. In \cite{nemirovski2007convex} and \cite{shang2020posteriori}, the Clopper-Pearson bound is presented to obtain less conservative conditional feasibility guarantees than with the Chernoff bound at the expense of increased computational complexity. For convex programs, \cite{shang2020posteriori} presents a secondary variational problem to use additional information from the attained solution, which outperforms the Clopper-Pearson bound for small sample sizes.
 
\textbf{Robust Approximations and Problems Beyond CCO.} Sampling-based solutions are not guaranteed to be feasible to \eqref{eq:cco} with probability one. In robust optimization, we construct (often overly large) uncertainty sets from convex approximations, e.g., using the conditional value at risk \cite{rockafellar2000optimization}, for computing solutions that always preserve the feasibility of  \eqref{eq:cco}. However,  these uncertainty sets usually have to be estimated from samples. Lastly, we mention work using  distributionally robust optimization, such as\cite{hota2019data}, to obtain a distributionally robust approximation of the CCO problem \eqref{eq:cco} by bounding the Wasserstein distance between an empirical distribution of samples $Y^{(1)}, \hdots, Y^{(K)}$ and $P_Y$. CCO problems have been extended to the distributionally robust setting where $P_Y$ is no longer fixed, but instead assumed to be contained within a set of distributions, also known as an ambiguity set. Solutions have been presented for ambiguity sets 
constructed with moments \cite{calafiore2006distributionally}, the Wasserstein Distance \cite{ji2021data}, or f-divergence measures \cite{jiang2016data}. Lastly, we remark on joint chance constrained optimization (JCCO) problems where multiple chance constraints are to be satisfied \cite{miller1965chance}. JCCO, inherently addressed by many of the aforementioned approaches (e.g., \cite{campi2008exact,luedtke2008sample}), can also be converted to CCO with multiple individual chance constraints using Boole's inequality (union bound) or pointwise maximum (both of which are methods standard in literature \cite{geng2019data}). 

In summary, compared with SA, CPP does not depend on support-constraint analysis, yielding dimension-independent certificates at the cost of additional calibration data. Compared with SAA, CPP adds an independent conformal prediction layer that provides finite-sample a posteriori certificates when standard SAA a priori guarantees are unavailable, difficult to compute, or lead to looser feasibility guarantees. Moreover, CPP is modular: by changing the conformal quantile, it can incorporate conformal variants for an extended class of CCO problems, as we show in this paper.

\section{Preliminaries}\label{sec:cp}

We next present conformal prediction and conditional conformal prediction, which we will use to solve CCO problems as in \eqref{eq:cco} with marginal and conditional feasibility guarantees, respectively.

\textbf{Conformal Prediction.} Conformal Prediction(CP) is a tool for uncertainty quantification that gained popularity for generating statistically valid prediction sets for machine learning models, see e.g., \cite{lindemann2025formal,angelopoulos2021gentle,angelopoulos2024theoretical} for an overview. Consider a set of independent and identically distributed (i.i.d.)\footnote{Conformal prediction  extends to exchangeable random variables, which is a weaker requirement than being i.i.d.} random variables $R, R^{(1)},\hdots,R^{(L)} \sim P_R$ where $P_R$ is an arbitrary distribution. One can think of $R$ as a test datapoint and of $R^{(1)},\hdots,R^{(L)}$ as a calibration dataset.  The variable $R$ is often referred to as the nonconformity score and can be the result of a function composition. For instance, in regression a common choice is the prediction error $R := |Z - \mu(U)|$ where $\mu$ is a predictor that predicts an output $Z$ from an input $U$.  CP now aims to find a probabilistic upper bound for $R$ based on $R^{(1)},\hdots,R^{(L)}$. The idea in CP is to compute the quantile over the empirical distribution of $R^{(1)},\hdots,R^{(L)}$  at a desired confidence level. Specifically, we define $\hat{Q}_{\alpha}(R^{(1)}, \hdots, R^{(L)}) \coloneqq \inf\{z \in \mathbb{R} \mid F_Z(z) \ge \alpha\}$ as the quantile at a confidence level $\alpha \in (0, 1)$ over the random variable $Z \coloneqq \sum_{i=1}^L\delta_{R^{(i)}}/L$ with $\delta_{R^{(i)}}$  being the unit point mass centered at $R^{(i)}$, and where $F_Z(\cdot)$ is the cumulative distribution function of $Z$. The next result summarizes the central idea behind CP where we denote by $\mathbb{P}_m(\cdot) \coloneqq \mathbb{P}^{L + 1}(\cdot)$ the product probability measure generated by the random variables $R,R^{(1)},\hdots,R^{(L)}$.\footnote{The subscript $m$ indicates ``marginal'' as is often referred to in the conformal prediction literature.} \\

\begin{lemma} \textbf{Quantile Lemma.} [Lemma 1 in \cite{tibshirani2019conformal}]\label{lemma:quantile_lemma}
    Let $R, R^{(1)}, \hdots, R^{(L)}$ be $L + 1$ i.i.d. random variables and $\delta \in (0, 1)$ be a failure probability so that $L \ge \lceil(L + 1)(1 - \delta)\rceil$. Then, it holds that
    \begin{align}\label{eq:cp_guarantee_marginal}
        \mathbb{P}_m(R \le C_m)\ge 1 - \delta,
    \end{align}
    where $C_m \coloneq \hat{Q}_{\alpha_m(L)}(R^{(1)}, \hdots, R^{(L)})$ is the quantile at confidence level $\alpha_m(L) \coloneqq (1 + 1/L)(1 - \delta)$. 
\end{lemma}
The quantile $C_m$ can be computed efficiently. Indeed, if $R^{(1)}, \hdots, R^{(L)}$ are sorted in nondecreasing order, it holds that $C_m = R^{(p)}$ where $p \coloneqq \lceil(L + 1)(1 - \delta)\rceil$, which makes it easy to compute the empirical quantile in practice, i.e., computing $C_m$ effectively reduces to computing the order statistics of $R^{(1)},\hdots,R^{(L)}$. 

The guarantees in \eqref{eq:cp_guarantee_marginal} are marginal over the randomness in $R, R^{(1)}, \hdots, R^{(L)}$, as indicated by the product measure $\mathbb{P}_m(\cdot)$. In other words, the statement in $\eqref{eq:cp_guarantee_marginal}$ is equivalent to $\mathbb{E}_{R^{(1)}, \hdots, R^{(L)}}[\mathbb{P}(R \le C_m)] \ge 1 - \delta$ using the total law of expectation, see \cite{angelopoulos2024theoretical} for details.  In fact, equation $\eqref{eq:cp_guarantee_marginal}$ will provide a less restrictive feasibility certificate (compared to those that generate conditional feasibility guarantees) to the CCO problem {under the relaxation via the product measure}, as we discuss further in Section \ref{sec:cpp}. We remark that $\mathbb{P}(R 
\ge C_m)$ is a random variable, whose distribution can be used for obtaining feasibility guarantees that are conditional on $R^{(1)},\hdots,R^{(L)}$. Such guarantees, however, are typically loose as we discuss in Section \ref{sec:appendix}\togglecomment{\cite{zhao2024conformal}}, motivating the next part of discussion.
 
\textbf{Conditional Conformal Prediction.} In CP, we obtained marginal guarantees for $R$ over the randomness in test and calibration data $R,R^{(1)},\hdots,R^{(L)}$ via the probability measure $\mathbb{P}_m(\cdot)$. In conditional CP\footnote{We are here interested in the training conditional variant in \cite{vovk2012conditional}. We drop the "training" term here for brevity.}, on the other hand, we obtain guarantees for $R$ that are, with high confidence, conditioned on the calibration data $R^{(1)},\hdots,R^{(L)}$. Interestingly, we can obtain such a guarantee using a tightened confidence level during the quantile computation. The next result summarizes the idea behind conditional CP  where we denote by $\mathbb{P}_L(\cdot) \coloneqq \mathbb{P}^L(\cdot)$ the product probability measure generated by the random variables $R^{(1)}, \hdots, R^{(L)}$.\\

\begin{lemma}\label{lemma:cond_quantile_lemma}
\textbf{Conditional Quantile Lemma.} [Proposition 2a in \cite{vovk2012conditional}] Let $R, R^{(1)}, \hdots, R^{(L)}$ be $L + 1$ i.i.d. random variables and $\delta\in(0,1)$ be a failure probability. Select $\beta \in (0, 1)$ and $1 - \beta$ be a confidence threshold such that $L \ge \lceil(L + 1)(1 - \delta + \sqrt{\frac{\ln(1/\beta)}{2L}})\rceil$. Then, it holds that $\mathbb{P}_L(\mathbb{P}(R \le C_c)\ge 1 - \delta) \ge 1 - \beta$, where $C_c \coloneq \hat{Q}_{\alpha_c(L)}(R^{(1)}, \hdots, R^{(L)})$ is the quantile at  confidence level $\alpha_c(L) \coloneqq (1 + 1/L)(1 - \delta + \sqrt{\frac{\ln(1/\beta)}{2L}})$.
\end{lemma}
Note that the quantile $C_c$ from Lemma \ref{lemma:cond_quantile_lemma} uses a more restrictive quantile level than the quantile $C_m$ from Lemma \ref{lemma:quantile_lemma}, i.e., that $C_c > C_m$. Therefore, $C_c$ also satisfies the marginal guarantee in equation \eqref{eq:cp_guarantee_marginal}, i.e. $\mathbb{P}_m(R \le C_c) \ge 1 - \delta$. We remark that \cite{vovk2012conditional} provides two other variants of conditional conformal prediction that, in some cases, provide less restrictive conditional bounds. We omit these variants for brevity, but note that CPP can similarly utilize these. We have so far assumed $R, R^{(1)}, \hdots, R^{(L)}$ to be identically distributed. If $R^{(1)}, \hdots, R^{(L)} \sim P_R$ and $R \sim P_{R_0}$, i.e., calibration and test data are from different distributions, then we need robust conformal prediction \cite{cauchois2024robust}. We summarize \cite{cauchois2024robust} in the appendix due to space limitation and rely on it to address robust CCO problems in Section \ref{sec:beyond_cco}.

\vspace{-7pt}
\section{Conformal Predictive Programming (CPP)}
\label{sec:cpp}
CPP consists of two main steps. In the optimization step, we approximate the optimization problem in \eqref{eq:cco} by replacing the chance constraint in \eqref{eq:chance_constraint_} with a  quantile constraint defined over an optimization dataset. We recall that all conformal prediction variants discussed in the previous section amount to computing quantiles at different confidence levels, hence allowing us to define different CPP variants thereby illustrating the versatility of CPP. We show that CPP can be viewed as an instantiation of SAA, and thus inherit a priori guarantees from the SAA literature for specific types of CCO problems. For general types of CCO problems, we introduce a calibration step, involving a second calibration dataset, to provide a posteriori feasibility guarantees.

\subsection{Chance Constrained Optimization via Quantile Reformulation} 
We present CPP for the two variants of CP presented in Section \ref{sec:cp}. Therefore, we select a quantile level of $\alpha(K)\in \{\alpha_m(K),\alpha_c(K)\}$. We next assume to have access to a dataset of $K$ i.i.d. random variables, which we refer to as the optimization dataset.\\
\begin{assumption}\label{assumption:cp}
   We have access to a dataset of $K$ i.i.d. random variables $Y^{(1)}, \hdots, Y^{(K)} \sim P_Y$ where $K$ is such that $K \ge \lceil (1 + K)(1 - \delta)\rceil$ if $\alpha(K)= \alpha_m(K)$ and $K \ge \lceil (1 + K)(1 - \delta+\sqrt{\frac{\ln{(1/\beta)}}{2K}})\rceil$ if $\alpha(K)= \alpha_c(K)$.\footnote{Note that $K \ge \lceil(K + 1)(1 - \delta)\rceil$ is equivalent to $K \ge \lceil 1/\delta\rceil -1$.}
\end{assumption}
We now consider Lemma \ref{lemma:quantile_lemma} to motivate CPP. For a fixed decision variable $x$ independent of test and optimization data $Y, Y^{(1)}, \hdots, Y^{(K)}$, we then directly know that $\mathbb{P}_m(f(x, Y)\le \hat{Q}_{\alpha_m(K)}(f(x, Y^{(1)}), \hdots,f(x, Y^{(K)})))\ge 1-\delta$. This motivates us, more generally, to approximate the optimization problem in \eqref{eq:cco} as
\begin{subequations}\label{eq:cp_quantile}
\begin{align}
    & \min_{x \in \mathcal{X}} \quad J(x)\\
    \textrm{s.t.} \ \ \ & \hat{Q}_{\alpha(K)}(f(x, Y^{(1)}), \hdots,f(x, Y^{(K)})) \!\le\! 0\label{eq:cp_quantile_cons}.
\end{align}
\end{subequations}
Due to the quantile constraint in \eqref{eq:cp_quantile_cons}, it is not immediately obvious how to solve the optimization problem \eqref{eq:cp_quantile}. Building up on existing work, we present three practical computational encodings of \eqref{eq:cp_quantile} in Section \ref{sec:appendix}. \togglecomment{the appendix  of \cite{zhao2024conformal}.} 

We denote the feasible region of the optimization problem in \eqref{eq:cp_quantile} as $F(K) \subseteq \mathbb{R}^n$. Note that the feasible region depends on $Y^{(1)},\hdots,Y^{(K)}$, which we indicate by the input argument $K$ in $F(K)$. Next, we denote the optimal solution by $x^*(K)$, again stressing the dependence on $Y^{(1)},\hdots,Y^{(K)}$. As the optimal solution $x^*(K)$ depends on $Y^{(1)}, \hdots, Y^{(K)}$, we  note that the random variables $f(x^*(K), Y), f(x^*(K), Y^{(1)}), \hdots, f(x^*(K), Y^{(K)})$ are no longer i.i.d. While $x^*(K)$ may be a feasible solution, this loss of independence means that we cannot apply Lemmas \ref{lemma:quantile_lemma} and \ref{lemma:cond_quantile_lemma} to make any formal statements about  $x^*(K)$. Following this observation, we first draw a connection with SAA in Remark \ref{remark:connection_saa} that enables us to obtain a priori guarantees of $x^*(K)$ for certain types of CCO problems. Afterwards, we discuss how to obtain a posteriori guarantees for general CCO problems.\\

\begin{remark}
\label{remark:connection_saa}
SAA approaches, such as in \cite{luedtke2008sample, luedtke2010integer, pagnoncelli2009sample}, approximate the CCO problem in \eqref{eq:cco} as
\begin{subequations}\label{eq:saa}
\begin{align}
    & \min_{x \in \mathcal{X}} \quad J(x)\\
    \textrm{s.t.} \ \ \ &\frac{1}{K}\sum_{i = 1}^K\mathbbm{1}\{f(x, Y^{(i)}) \le 0\} \ge 1 - \omega\label{eq:ssa_cons}.
\end{align}
\end{subequations}
\noindent where $\omega \in (0, 1)$ is a user-defined parameter.  For any $\omega\in (0,\delta)$, SAA provides a priori conditional feasibility guarantees if (1) $\mathcal{X}$ is finite, (2) the chance constraint \eqref{eq:chance_constraint_} is of separable form $f(x, Y) = Y - g(x)$ for some  function $g:\mathbb{R}^n\to \mathbb{R}$, or (3) $f$ is Lipschitz continuous. SAA also provides a priori optimality guarantees. We summarize the feasibility and optimality guarantees in the Appendix in Section \ref{sec:appendix}\togglecomment{appendix of \cite{zhao2024conformal}}. We remark that these guarantees can be loose, further motivating a posteriori guarantees.  Next, note that
\begin{align*}
\begin{split}
    \eqref{eq:cp_quantile_cons} \Leftrightarrow
    \begin{cases}
        \sum_{i = 1}^K&\mathbbm{1}\{f(x, Y^{(i)}) \le 0\} \ge \\
        &\lceil(K + 1)(1-\delta)\rceil \text{ if } \alpha(K)=\alpha_m(K)\\
        \sum_{i = 1}^K&\mathbbm{1}\{f(x, Y^{(i)}) \le 0\} \ge \\
        &\hspace{-1.2cm}\lceil(K + 1)(1-\delta+\sqrt{\frac{\ln{(1/\beta)}}{2K}})\rceil \text{ if } \alpha(K)=\alpha_c(K)
    \end{cases}
    \end{split}
\end{align*}
This means that \eqref{eq:cp_quantile_cons} is equivalent to \eqref{eq:ssa_cons} if (1) $\omega := 1 - \frac{\lceil(K + 1)(1-\delta)\rceil}{K}$ for $\alpha(K)=\alpha_m(K)$, and (2) $\omega := 1 - \frac{\lceil(K + 1)(1-\delta+\sqrt{\frac{\ln{(1/\beta)}}{2K}})\rceil}{K}$ for $\alpha(K)=\alpha_c(K)$, allowing us to obtain a priori guarantees for CPP from SAA. In CPP, $\omega$ is chosen to match the conformal quantile rule from the subsequent calibration step, thereby aligning the optimization problem with the certificate that will be computed from the calibration data; this same mechanism allows different conformal variants, such as robust conformal prediction, to be incorporated by changing the corresponding quantile rule, see examples in Section \ref{sec:beyond_cco}.

\end{remark}

Besides Remark \ref{remark:connection_saa}, we can directly obtain a priori feasibility guarantees for separable constraints.\\
\begin{lemma}
\label{lemma:separable}
    Suppose the  function $f(x, Y)$ is of the form $f(x, Y) \coloneq h(Y) - g(x)$ where $h:\mathbb{R}^d \rightarrow \mathbb{R}$ and $g:\mathbb{R}^n \rightarrow \mathbb{R}$ are arbitrary functions. Then, it holds that $\mathbb{P}^{K + 1}(f(x^*{(K)}, Y) \le 0) \ge 1 - \delta$, where $x^*(K)$ can be  any feasible solution to \eqref{eq:cp_quantile} for $\alpha(K)=\alpha_m(K)$. Furthermore, it holds that, $\mathbb{P}^{K}(\mathbb{P}(f(x^*{(K)}, Y) \le 0) \ge 1 - \delta)\ge 1-\beta$, where $x^*(K)$ can be  any feasible solution to \eqref{eq:cp_quantile} for $\alpha(K)=\alpha_c(K)$. 
\end{lemma}


\subsection{A posteriori Feasibility Guarantees via Calibration}
As said, we generally cannot obtain a priori guarantees since $f(x^*(K), Y), f(x^*(K), Y^{(1)}),\hdots, f(x^*(K), Y^{(K)})$ are not independent anymore as $x^*(K)$ was trained on the optimization dataset $Y^{(1)},\hdots,Y^{(K)}$. To obtain a posteriori guarantees, we need a second independent dataset, which we refer to as the calibration dataset.\\

\begin{assumption}\label{assumption:cp_post}
    We have access to a dataset  of $L$ i.i.d. random variables $Y^{(K + 1)}, \hdots, Y^{(K + L)} \sim P_Y$ such that $L \ge \lceil (1 + L)(1 - \delta)\rceil$ if $\alpha(L)= \alpha_m(L)$ and $L \ge \lceil (1 + L)(1 - \delta+\sqrt{\frac{\ln{(1/\beta)}}{2L}})\rceil$ if $\alpha(L)= \alpha_c(L)$.
\end{assumption}

We will now provide marginal and conditional a posteriori feasibility guarantees.

\textbf{Marginal Feasibility Guarantees.} We now calibrate the solution $x^*(K)$ using the calibration dataset and Lemma \ref{lemma:quantile_lemma} to obtain marginal feasibility guarantees. In essence, we perform a conformal prediction step with the nonconformity score 
\begin{align*}
  R^{(i)} \coloneq f(x^*(K), Y^{(i)}) \text{ for } i \in \{K + 1, \hdots, K + L\}   
\end{align*}
for which we compute the quantile $C_m(x^*(K)) \!\coloneqq\! \hat{Q}_{\alpha_m(L)}(f(x^*(K), Y^{(K + 1)}), \hdots, f(x^*(K), Y^{(K + L)}))$ so that $C_m(x^*(K))$ is a probabilistically valid upper bound on $f(x^*(K), Y)$, as summarized next.\\

\begin{theorem}\label{thm:a_posteriori_marginal}\textbf{Marginal Guarantees.} Let Assumption \ref{assumption:cp_post} hold. Then, the solution $x^*(K)$ of the CPP problem \eqref{eq:cp_quantile} with $\alpha(K)=\alpha_m(K)$ is such that  $x^*(K) \in \mathcal{X}$ and $\mathbb{P}_m(f(x^*(K), Y) \le C_m(x^*(K))) \ge 1 - \delta.$
\end{theorem}

In the previous subsection, we considered  Assumption \ref{assumption:cp}, i.e.,  we assumed that the optimization dataset  $Y^{(1)}, \hdots, Y^{(K)}$ was i.i.d. This allowed us to obtain a priori feasibility guarantees, e.g., as in Remark \ref{remark:connection_saa}. However, the optimization dataset is not required to be i.i.d. in Theorem \ref{thm:a_posteriori_marginal}. Nonetheless, it is clear that selecting an optimization dataset that is not i.i.d. can result in unnecessarily large upper bounds $C_m(x^*(K)))$. Therefore, it is generally recommended that both Assumptions \ref{assumption:cp} and \ref{assumption:cp_post} hold.\\

\begin{remark}\label{remark:marginal}
As evident from the proof of Theorem \ref{thm:a_posteriori_marginal}, the guarantees in Theorem \ref{thm:a_posteriori_marginal}  hold for any feasible solution $x$ of the CPP problem \eqref{eq:cp_quantile}. Additionally, choosing a quantile level in \eqref{eq:cp_quantile} that is different from $\alpha(K)=\alpha_m(K)$ does not affect the validity of Theorem \ref{thm:a_posteriori_marginal}. However, this may once again  result in unnecessarily large upper bounds $C_m(x^*(K))$, while we ideally want that $C_m(x^*(K)) \le 0$ to approximate the original chance constraint \eqref{eq:chance_constraint_}. By satisfying Assumption \ref{assumption:cp} and by selecting $\alpha(K)=\alpha_m(K)$, the empirical quantile enforced during optimization is aligned with the empirical quantile later evaluated during calibration. When the objective and chance constraint are in tension, the quantile constraint is typically active or nearly active at the optimizer; hence, if the optimization and calibration datasets have comparable finite-sample behavior, we expect $C_m(x^*(K))$ to be small or close to zero in practice. We will empirically demonstrate this behavior in our experiments in Section \ref{sec:case_studies} and, in the next section, introduce the idea of a quantile shift to provide a posteriori conditional feasibility guarantees for $f(x^*(K), Y) \le 0$.
\end{remark}

\textbf{Conditional Feasibility Guarantees.} We now aim to provide conditional guarantees. We again calibrate the solution $x^*(K)$, but now via  Lemma \ref{lemma:cond_quantile_lemma} to obtain conditional a posteriori feasibility guarantees. We compute the quantile $C_c(x^*(K)) \!\coloneqq\! \hat{Q}_{\alpha_c(L)}(f(x^*(K), Y^{(K + 1)}), \hdots
    , f(x^*(K), Y^{(K + L)}))$, and obtain the following result, where the proof is omitted as it similarly follows Theorem \ref{thm:a_posteriori_marginal}, but now with Lemma \ref{lemma:cond_quantile_lemma} instead of Lemma \ref{lemma:quantile_lemma}.\\

\begin{theorem}\label{thm:a_posteriori_conditional}\textbf{Conditional Guarantees.} Let Assumption \ref{assumption:cp_post} hold. Then, the solution $x^*(K)$ of the CPP problem \eqref{eq:cp_quantile} with $\alpha(K)=\alpha_c(K)$ is such that  $x^*(K) \in \mathcal{X}$ and $\mathbb{P}_L(\mathbb{P}(f(x^*(K), Y) \le C_c(x^*(K)) \ge 1 - \delta) \ge 1 - \beta$.
\end{theorem}
As in Remark \ref{remark:marginal}, note that Theorem \ref{thm:a_posteriori_conditional} holds for any feasible solution $x$ of the CPP problem \eqref{eq:cp_quantile}. At the same time, selecting $\alpha(K)=\alpha_c(K)$ in \eqref{eq:cp_quantile} aligns the optimization and calibration quantile rules, so we expect $C_c(x^*(K))$ to be small or close to zero in practice. As discussed earlier, this bound also provides a valid but generally more restrictive marginal feasibility guarantee $\mathbb{P}_m(f(x^*(K), Y) \le C_c(x^*(K))) \ge 1 - \delta$. 

In Theorems \ref{thm:a_posteriori_marginal} and \ref{thm:a_posteriori_conditional}, we note that we cannot ensure nonpositivity of $C_m(x^*(K))$ and $C_c(x^*(K))$, respectively. To address this issue, we can instead compute the failure probability $\delta^*$ that guarantees $C_c(x^*(K)) \le 0$. \\

\begin{theorem}\label{thm:quantile_shift}
    \textbf{Quantile Shift.} Let Assumption \ref{assumption:cp_post} hold. Define the adjusted probability $\delta^* \coloneq 1 - \frac{S}{L + 1} + \sqrt{\frac{\ln(1/\beta)}{2L}}$ where $S \coloneq \sum_{i = K+1}^{K+L}\mathbbm{1}(f(x^*(K), Y^{(i)}) \le 0)$ is the satisfaction count of the constraint $f$ under $x^*(K)$. If $\delta^* \in (0, 1)$, the solution $x^*(K)$ of the CPP problem \eqref{eq:cp_quantile} with $\alpha(K)=\alpha_c(K)$ is such that  $x^*(K) \in \mathcal{X}$ and
    \begin{align}\label{eq:shift_cond_2}
        \mathbb{P}_L(\mathbb{P}(f(x^*{(K)}, Y) \le 0) \ge 1 - \delta^*) \ge 1 - \beta.
    \end{align}
\end{theorem}
\vspace{-16pt}

We emphasize that Theorem \ref{thm:quantile_shift} hinges on the assumption that $\delta^* \in (0, 1)$, and that $\delta^*$ itself is a random variable as it depends on the calibration dataset. As before, Theorem \ref{thm:quantile_shift} is valid for any feasible solution $x$ of the CPP Problem \eqref{eq:cp_quantile}. We also remark that $S$ can efficiently be computed in linear time. This is in contrast to, for instance, the scenario optimization results from \cite{campi2018general, garatti2024non} for nonconvex CCOs, as we further empirically compare in Section \ref{sec:case_studies}. Interestingly, we can show that the quantile shift result in equation \eqref{eq:shift_cond_2} effectively reduces to the one-sided Chernoff bound from \cite{shang2020posteriori} for large calibration datasets, i.e., for large $L$ (see the appendix \togglecomment{of \cite{zhao2024conformal}}in Section \ref{sec:appendix}). We can, in the same way, derive similar quantile shift results using the other two variants of conditional CP, as presented in \cite{vovk2012conditional}. As discussed in Section \ref{sec:cp}, these variants provide advantages for certain ranges of $L$.

\section{Beyond Standard CCO Problems}
\label{sec:beyond_cco}
CP has been an active research area with developments in adaptive CP \cite{gibbs2021adaptive, zaffran2022adaptive}, robust CP \cite{cauchois2024robust, gendler2021adversarially}, conformalized quantile regression \cite{romano2019conformalized, sesia2020comparison}, outlier detection \cite{lei2015conformal, guan2022prediction}, Mondrian CP \cite{bostrom2020mondrian, angelopoulos2021gentle, gibbs2023conformal, ding2024class}, and many other variants. The key observation here is that these variants always rely on computing an empirical quantile, and that they only differ in the choice of the nonconformity score and the quantile level. We argue that the strength of the CPP framework is that it can easily be generalized to incorporate different variants of CP. To illustrate this, we present robust conformal predictive programming (RCPP) and propose Mondrian CCO, solved with Mondrian CPP.

\textbf{Robust Conformal Predictive Programming (RCPP).} RCPP can deal with distribution shifts in $P_Y$, i.e., when the datapoint $Y$ is not following the distribution $P_Y$ from which the optimization and calibration datasets are drawn. This may be the case when optimization and deployment conditions are different, e.g. when there is a sim2real gap as often is the case in robotics application.  RCPP is based on robust conformal prediction as presented  in Section \ref{sec:cp}. We assume that $Y$ now follows a distribution from the ambiguity set $\mathcal{P}(P_Y, \epsilon) \coloneqq \{P_{\tilde{Y}} \mid D_\phi(P_{\tilde{Y}}, P_Y) \le \epsilon\}$ where $\epsilon>0$ is a parameter chosen a priori (which we denote as the distribution shift) and $D_\phi$ is an f-divergence measure. In essence, robust CP follows the same procedure as CP but uses a tightened quantile level  $\alpha_r(K)$ such that $\alpha_r(K)>1-\delta$. We note that $\alpha_r(K)$ and an auxiliary function $v$ are defined in the \togglecomment{appendix of \cite{zhao2024conformal}}Appendix in Section \ref{sec:appendix}.

We demonstrate the use of robust CP in solving robust chance constraint optimization (RCCO) of the form
\begin{subequations}\label{eq:rcco}
\begin{align}
\min_{x \in \mathcal{X}} \quad & J(x)\\
\textrm{s.t.} \quad & \inf_{Y \sim P_{Y'} \in \mathcal{P}(P_Y, \epsilon)} \mathbb{P}(f(x, Y) \le 0) \ge 1 - \delta.
\end{align}
\end{subequations}
The difference between the RCCO in \eqref{eq:rcco} and the CCO in \eqref{eq:cco} is that $Y$ is no longer drawn from the distribution $P_Y$, but is instead drawn from a distribution $P_{Y'}$ within the ambiguity set $\mathcal{P}(P_Y, \epsilon)$. Our robust extension of CPP uses robust CP and requires the next assumption.\\

\begin{assumption}\label{assumption:drcpp}
   We make the same assumptions on optimization and calibration datasets as in Assumption \ref{assumption:cp} and \ref{assumption:cp_post}, but require now that  $K, L \ge \Big\lceil \frac{v^{-1}(1 - \delta)}{1 - v^{-1}(1 - \delta)} \Big\rceil$.
\end{assumption}


Similar to CPP, RCPP consists of a quantile reformulation for optimization and an a posteriori feasibility analysis, which we summarize in the following theorem.\\

\begin{theorem}\label{thm:rcpp_marginal}
     Let Assumption \ref{assumption:drcpp} hold. Then, the solution $x^*(K)$ of the CPP problem \eqref{eq:cp_quantile} with the quantile level $\alpha(K)=\alpha_r(K)$ is such that $x^*(K) \in \mathcal{X}$ and $  \inf_{Y \sim P_{Y'} \in \mathcal{P}(P_Y, \epsilon)} \tilde{\mathbb{P}}_m(f(x^*(K), Y) \le \tilde{C}(x^*(K))) \ge 1 - \delta,$ where $\tilde{C}(x^*(K)) \coloneq \hat{Q}_{\alpha_r(L)}(f(x^*(K), Y^{(K + 1)}), \hdots, \\f(x^*(K), Y^{(K + L)}))$ and $\tilde{\mathbb{P}}_m \coloneq \mathbb{P}_Y^K \times \mathbb{P}_{Y'}$.
\end{theorem}

\textbf{Mondrian Conformal Predictive Programming (MCPP).} In MCPP, we deal with chance constraints that are conditioned on $Y$ belonging to a certain class. As a motivating example, consider the problem of synthesizing an optimal motion plan $x^*$ for a robot under stochastic sensor noise $Y$. We want to ensure that $f(x^*, Y) \le 0$ with probability no less than $1-\delta$, but not over the distribution of $P_Y$ and instead over the distribution of $P_Y$ conditioned on $Y$ belonging to a specific range. One specific instance could be when $Y$ is Gaussian distributed and we want to verify that $\mathbb{P}(f(x^*, Y) \le 0 \mid Y \in G) \ge 1 - \delta$ for all ranges $ G \in \mathcal{G}$ where $\mathcal{G} \coloneq \{(-\infty, -0.1), [-0.1, 0.1], (0.1, \infty)\}$. This allows us to reason over the policy $x^*$ in ensuring safety against high-likelihood and low-likelihood events. 

This motivates us to define the problem of MCCO as
\begin{subequations}\label{eq:mcco}
\begin{align}
\min_{x \in \mathcal{X}} \quad & J(x)\\
\textrm{s.t.} \quad & \mathbb{P}(f(x, Y) \le 0 \mid Y \in G) \ge 1 - \delta, \forall G \in \mathcal{G}, \end{align}
\end{subequations}
where $\mathcal{G}$ is a user defined set of classes and $\cup_{G \in \mathcal{G}}G \subseteq \Xi$ with $\Xi \subseteq \mathbb{R}^d$ denoting the support of $Y$. Our goal is to synthesize a single solution $x^*$ that is valid for all classes $G \in \mathcal{G}$, while the group of $Y$ is not known a priori. With the assumption of a priori lack of knowledge of $Y$, one cannot simply apply CPP to attain different solutions to different groups (which would also be intractable when the number of groups $|\mathcal{G}|$ is large). To solve the MCCO \eqref{eq:mcco}, we propose MCPP where we compute a feasible solution $x^* \coloneq x^*(K)$ of the CPP Problem \eqref{eq:cp_quantile}, but then perform a modified calibration step for obtaining a posteriori feasibility guarantees. Our approach is motivated by Mondrian CP \cite{angelopoulos2021gentle}.\footnote{We are motivated by the class-conditional conformal prediction from \cite{angelopoulos2021gentle}, but instead focus on instances where $Y\in G$.}  For simplicity, we focus on marginal guarantees via $\alpha=\alpha_m$, while the extension to conditional guarantees via $\alpha=\alpha_c$ is straightforward. We also omit the proof since it follows similarly to before from Lemmas \ref{lemma:quantile_lemma} and \ref{lemma:cond_quantile_lemma}.\\
\begin{theorem}
    \label{thm:classwise}  Consider a set-valued function $\Gamma$ that maps a group $G$ to a set of samples such that $\Gamma(G) \coloneq \{Y^{(i)} \mid  Y^{(i)} \in G \text{ for } i \in \{K + 1, \hdots, K + L\}\}$. Suppose $|\Gamma(G)| \ge \lceil(|\Gamma(G)| + 1)(1 - \delta)\rceil$ for all $G \in \mathcal{G}$. Then, for all $ G \in \mathcal{G}$, the solution $x^*(K)$ of the CPP problem \eqref{eq:cp_quantile} with $\alpha(K)=\alpha_m(K)$ is such that $x^*(K) \in \mathcal{X}$ and $\mathbb{P}^{|\Gamma(G)| + 1}(f(x^*(K), Y) \le C_G \mid Y \in G) \ge 1 - \delta,$ where $C_G \coloneq \hat{Q}_{\alpha_m(|\Gamma(G)|)}(\{f(x^*(K), Y^{(i)}) \mid Y^{(i)} \in \Gamma(G)\})$.
\end{theorem}

\section{Case Studies}\label{sec:case}
\label{sec:case_studies}
We validate CPP on convex and nonconvex CCO case studies, showing its advantage over SA \cite{campi2018general, garatti2024non} in the nonconvex setting. We also evaluate RCPP and MCPP on a stochastic optimal control problem, starting with an overview of the experimental setup.
\begin{figure*}
    \centering
    \includegraphics[width = 0.75\linewidth]{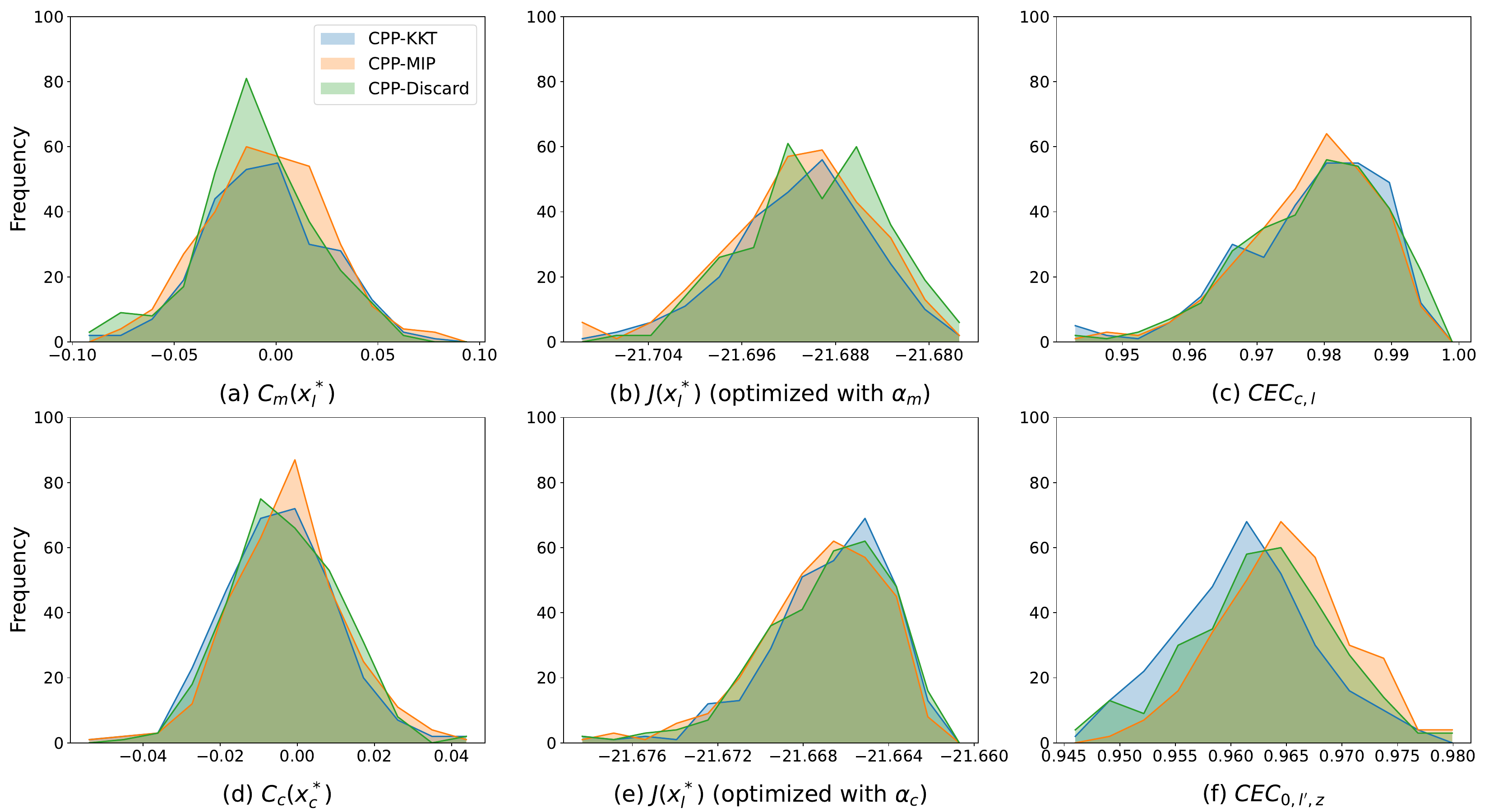}
    \caption{Results for Section \ref{sec:convex_case} (Convex Problem)}
    \label{fig:case1}
\end{figure*}

\textbf{Experimental Procedure.} In each of our case studies, we a priori choose parameters $\delta$ and $\beta$. We let $N$ denote the number of experiments, and $K$ and $L$ again denote the size of optimization and calibration datasets. Specifically, we perform the following procedure for computational encodings of CPP-Bilevel, CPP-MIP, and CPP-Discarding as described in\togglecomment{the appendix \cite{zhao2024conformal}} the Appendix in Section \ref{sec:appendix}. For each experiment $l \in \{1, \hdots, N\}$, we sample an optimization dataset $Y_l^{(1)}, \hdots, Y_l^{(K)} \sim P_Y$ where $P_Y$ is problem specific. We then compute the solution $x^*_l$ of $\eqref{eq:cp_quantile}$ with $\alpha_m(K), \alpha_c(K)$, or $\alpha_r(K)$ depending on the guarantee to be evaluated.

\textbf{Evaluating  Marginal Feasibility Guarantees.} In each experiment $l \in \{1, \hdots, N\}$, we sample a calibration dataset $Y_l^{(K + 1)}, \hdots, Y_l^{(K + L)} \sim P_Y$. We then compute the upper bound $C_m(x^*_l)$ (which we replace with other variants for Mondrian CPP and RCPP) according to Theorem \ref{thm:a_posteriori_marginal}. At the end of $N$ experiments, we compute the empirical coverage of the solution with respect to $C_m(x^*_l)$, where $EC \coloneq \frac{1}{N}\sum_{l = 1}^N\mathbbm{1}(f(x^*_l, Y_l^{(K + 1)}) \le C_m(x^*_l)).$ As $N$ approaches $\infty$, we expect (and should observe) $EC$ to converge to a value larger than $1 - \delta$ according to Theorem \ref{thm:a_posteriori_marginal}. We also show the histograms of $C_m(x_l^*)$ and $J(x_l^*)$ across the $N$ experiments. When evaluating Mondrian CPP, we additionally evaluate the Mondrian empirical coverage, which we denote by $MEC(C, G)$ where $G \subseteq \mathbb{R}^d$ is an a priori determined test group and $C$ can be $C_m$ or $C_G$. To find $MEC(C, G)$, we evaluate $EC$ but simultaneously require that $Y_l^{(K + 1)} \sim P_Y$ belongs to an a priori determined test group $G$ for each experiment $l$. Note that if $C \coloneq C_G$, we expect $MEC(C, G)$ to converge to a value greater than $1 - \delta$ if $\Gamma(G)$ holds consistent over the experimental trials and if $N$ approaches $\infty$. Since we cannot control $|\Gamma(G)|$ in practice for each calibration set, we emphasize $MEC(C, G)$ is only an empirical estimation on the coverage guarantee in Theorem \ref{thm:classwise}. As a comparison, we record $MEC(C_m, G)$, which we do not expect to achieve $1 - \delta$ coverage. 

\begin{figure*}
    \centering
    \includegraphics[width = 0.75\linewidth]{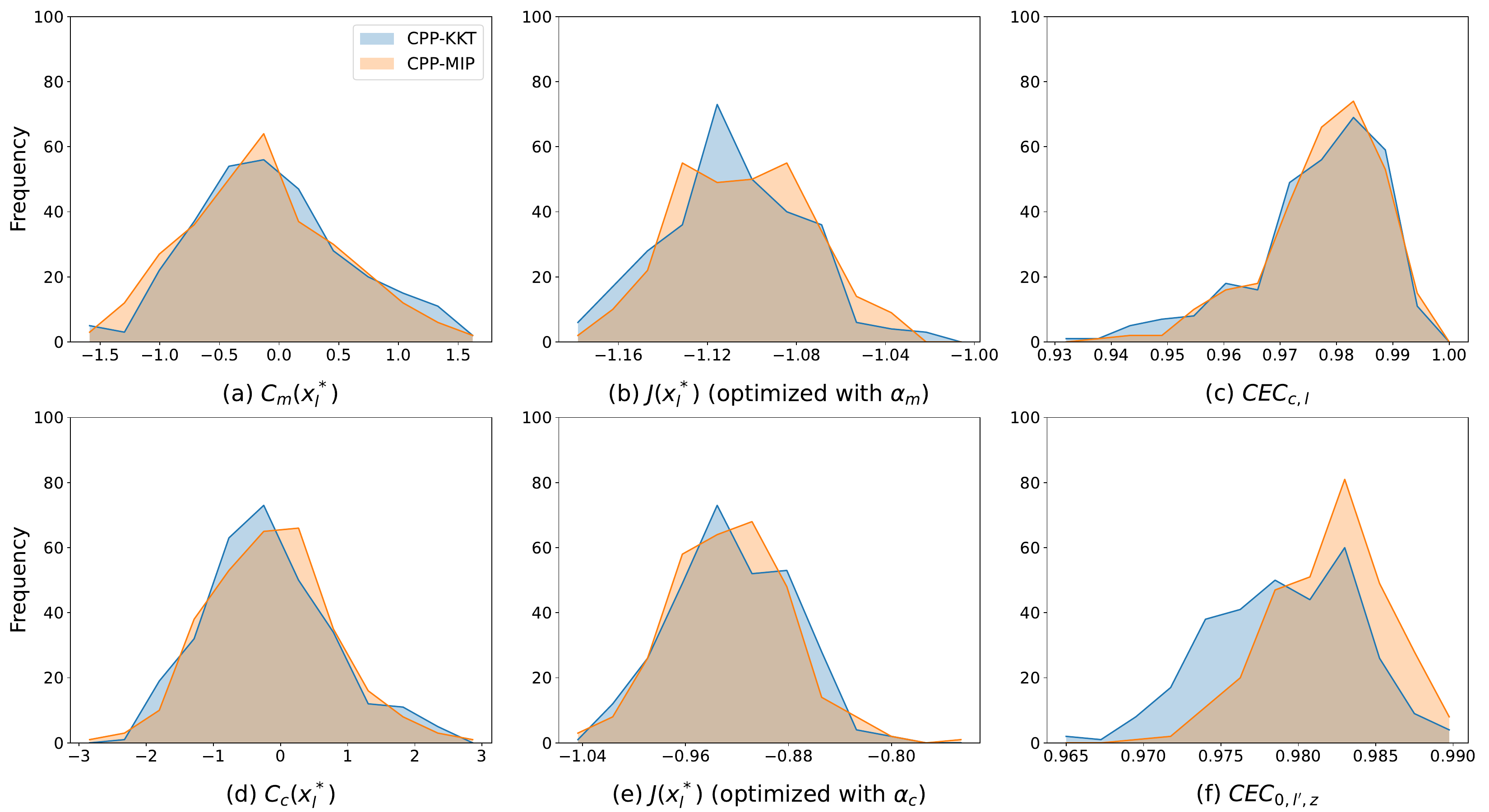}
    \caption{Results for Section \ref{sec:nonconvex_case} (Nonconvex Problem)}
    \label{fig:case2}
\end{figure*}

\textbf{Evaluating  Conditional Feasibility Guarantees.} In each experiment, we again sample a calibration dataset $Y_l^{(K + 1)}, \hdots, Y_l^{(K + L)} \sim P_Y$, but now compute $C_c(x_l^*)$ according to Theorem \ref{thm:a_posteriori_conditional}. In each experiment, we  additionally sample $V$ independent test datapoints $Y_l^{(K + L + 1)}, \hdots, Y_l^{(K + L + V)} \sim P_Y$ and compute the conditional empirical coverage of the solution with respect to $C_c(x_l^*)$, where $CEC_{C, l} \coloneq \frac{1}{V}\sum_{i = K + L + 1}^{K + L + V}\mathbbm{1}(f(x_l^*, Y_l^{(i)}) \le C_c(x_l^*))$. At the end of the experiment we plot the histograms of $CEC_{C, l}$, $C_c(x_l^*)$ and $J(x_l^*)$ across the $N$ experiments. As $N$ and $V$ approach $\infty$, we should expect the histogram of $CEC_{C, l}$ to approximate the shape of the probability density function of $\mathbb{P}(f(x_l^*, Y) \le C_c(x_l^*))$ for which we know that $\mathbb{P}(f(x_l^*, Y) \le C_c(x_l^*)) \ge 1 - \delta$ with a probability of no less than $1 - \beta$. To evaluate the quantile shift, we choose one experiment $l' \coloneq 1$ from the experiments and calculate $\delta^*_{l'}$ following Theorem \ref{thm:quantile_shift}. We then draw $Z$ sets of $W$ samples $Y_{l', z}^{(K + L + V + 1)}, \hdots Y_{l', z}^{(K + L + V + W)} \sim P_Y$ for $z \in \{1, \hdots, Z\}$. For $z \in \{1, \hdots, Z\}$, we compute $CEC_{0, l', z} \coloneq \frac{1}{W}\sum_{i = K + L + V + 1}^{K + L + V + W}\mathbbm{1}(f(x_{l'}^*, Y_{l', z}^{(i)}) \le 0)$. We then plot the histogram of $CEC_{0, l', z}$ and expect it to approximate the shape of the probability density function of $\mathbb{P}(f(x_{l'}^*, Y) \le 0)$ as $Z$ and $W$ approach $\infty$. We know (and should observe) that  $\mathbb{P}(f(x_{l'}^*, Y) \le 0)\ge 1 - \delta^*_{l'}$ with a probability no less than~$1 - \beta$. Computation of $x_l^*$ is conducted with the SCIP optimization solver \cite{achterberg2009scip}, and $x_l^*$ and $\delta^*$ are computed on a MacBook Air with Apple M2 and 16 GB of RAM. We disregard any solution obtained after 200 seconds (timeout) and any infeasible solution.

\subsection{Numerical Case Studies}

\subsubsection{Numerical Case Study with a Convex Problem}\label{sec:convex_case}

\textbf{Problem Statement.} We consider the  CCO problem 
\begin{align}\label{eq:casestudy_convex}
    \!\!\! \min_{x \in \mathbb{R}^2} c^\top x \ \
    \textrm{s.t.}  \text{Prob}((x_1\!-\!3)^2 + (x_2\!-\!5)^2 \!\le\! Y ) \!\ge\! 1 \!-\! \delta,
\end{align}
where $x \coloneqq [x_1, x_2]^\top$ is a 2-dimensional variable, $c \coloneqq [-1, -2]^\top$ is a vector. Note that  cost and  constraint functions are  convex.
The failure probability is set to $\delta \coloneqq 0.1$ and $Y \sim \mathcal{U}(15, 16)$ follows a uniform distribution. 

\noindent \textbf{Results.} For now, we fix $K := 200$, $L := 200$, $\beta := 0.1$, $N := 300$, $V := 1000$, $Z := 300$, $W := 1000$.
We conduct both marginal and conditional validation as described in the previous subsection using the three proposed encoding methods. We observe an $EC$ of $0.91$, $0.87$, and $0.92$ respectively for CPP-KKT, CPP-MIP and CPP-Discarding. The resulting plots are presented in Fig. \ref{fig:case1}. In the marginal case, the empirical results for $C_m(x^*_l)$ and $J(x^*_l)$ are shown in Fig. \ref{fig:case1}(a) and \ref{fig:case1}(b), respectively. For the conditional case, $CEC_{C, l}$, $C_c(x^*_l)$, and $J(x^*_l)$ are illustrated in Fig. \ref{fig:case1}(c), \ref{fig:case1}(d), and \ref{fig:case1}(e). As expected, both histograms of $C_m(x^*_l)$ and $C_c(x^*_l)$ center near 0 and $J(x^*_l)$ is larger when optimized with $\alpha_c$ as compared to with $\alpha_m$. Regarding the quantile shift, the $\delta^*_{l'}$ values for the first experiment $l'=1$ across the three encoding methods CPP-KKT, CPP-MIP, and CPP-Discarding are 0.12, 0.13 and 0.11, respectively. Additionally, $CEC_{0, l', z}$ is shown in Fig. \ref{fig:case1}(f).
\begin{figure*}
    \centering
    \includegraphics[width = 0.75\linewidth]{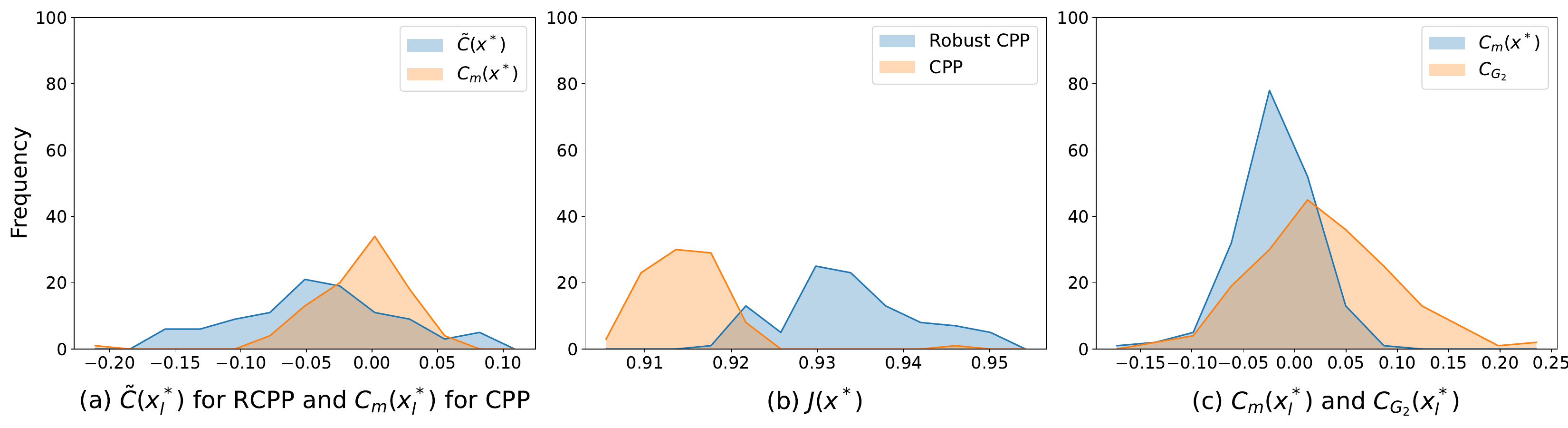}
    \caption{Results for Section \ref{sec:stochastic} (Stochastic Optimal Control)}
    \label{fig:case3}
\end{figure*}

\subsubsection{Numerical Case Study with a Nonconvex Problem}\label{sec:nonconvex_case}

\textbf{Problem Statement.} We consider the CCO problem 
\begin{align}\label{eq:casestudy_nonconvex}
    \min_{x \in \mathbb{R}} \quad x^3e^x \quad \text{ s.t.} \quad & \text{Prob}(50Ye^{x} - 5\le 0) \ge 1 - \delta, \nonumber \\
    \quad &  x^3 + 20 \le 0
\end{align}
with failure probability $\delta \coloneqq 0.1$ and where $Y \sim \text{Exp}(\frac{1}{3})$ is a long-tailed exponential distribution. We emphasize that this CCO problem is in the nonconvex setting.

\noindent \textbf{Results.} For now, we again fix $K := 200$, $L := 200$, $\beta := 0.1$, $N := 300$, $V := 1000$, $Z := 300$, $W := 1000$.
We conduct the same experiment as before using two encoding methods, CPP-KKT and CPP-MIP. We do not compare to CPP-Discard since it relies on the convexity assumption. We observe an EC of 0.90 and 0.89 respectively for CPP-KKT and CPP-MIP. The resulting plots are presented in Fig. \ref{fig:case2}. We again observe that histograms of $C_m(x_l^*)$ and $C_c(x_l^*)$ center near 0 and the problem is more costly when optimized with $\alpha_c$ than with $\alpha_m$.

We compare the computational complexity of our methods with those proposed in nonconvex SA \cite{campi2018general, garatti2024non} via finding the irreducible support subsample following \cite{campi2018general}. Although the average computation time to solve for $x_l^*$ with nonconvex SA in \cite{campi2018general} and \cite{garatti2024non} is 0.06 seconds\footnote{When optimzed with $\alpha_m$, we observe an average computation time of 6.21 seconds for CPP-KKT and an average computation time of 0.20 for CPP-MIP. When optimized with $\alpha_c$, we observe an average computation time of 4.37 seconds for CPP-KKT and an average computation time of 0.18 seconds for CPP-MIP.}, we observe that the average computation times for $\delta^*$ in \cite{campi2018general} and \cite{garatti2024non} are 10.26 and 10.29 seconds respectively, which are greater than our computation times. 



\subsection{Stochastic Optimal Control}\label{sec:stochastic}
We demonstrate the effectiveness and utility of RCPP and Mondrian CPP in solving a stochastic optimal control problem.
 Consider a robot operating in a two-dimensional Euclidean space, e.g, a mobile service robot. The state of the robot is $y_t \coloneqq [x^{(1)}_t, v^{(1)}_t, x^{(2)}_t, v^{(2)}_t] \in \mathbb{R}^4$ where $x^{(1)}_t, v^{(1)}_t$ and $x^{(2)}_t, v^{(2)}_t$ represent position and velocity at time $t$ in each dimension. We describe the robot dynamics by discrete-time double integrator dynamics $y_{t + 1} = Ay_t + Bu_t + w_t, \; y_0 := (0, 0, 0, 0)^T$ with $ A \coloneqq ((1, 1, 0, 0), (0, 1, 0, 0), (0, 0, 1, 1), (0, 0, 0, 1))$, $ 
    B \coloneqq ((0.5, 0), (1, 0), (0, 0.5), (0, 1))$ and where $u_t \in \mathbb{R}^2$ is the input and $w_t \in \mathbb{R}^4$ is system noise sampled from a predefined distribution (described later). We let $T \coloneqq 5$ be a user-specified time horizon and denote the multivariate system noise by $w \coloneq (w_0, \hdots, 
    w_{T - 1}) \in \mathbb{R}^{T \times 4}$. We are interested in synthesizing control inputs $u_t$ for times $t = 0, \hdots, T - 1$ that allow the robot to reach a circle centered around the target location $[5, 5]$ at time $T$ with high probability. Specifically, for parameters $\delta \coloneqq 0.1$ and $\zeta \coloneqq 1$, we want to solve the control problem
\begin{align*}
    \min_{u} \ \sum_{t = 0}^{T} &   \ \ \ \|u_t\|_2^2 \quad
    \textrm{s.t.} \quad
    y_0 = (0, 0, 0, 0)^T,\nonumber\\
    & y_{t + 1} = Ay_t + Bu_t + w_t, \forall t \in \{1, \hdots, T-1\} \nonumber, \\
    & \mathbb{P}((x_T^{(1)} - 5)^2 + (x_T^{(2)} - 5)^2\le \zeta)  \ge 1 - \delta. \nonumber
\end{align*}
We first evaluate RCPP with a Robust CCO.

\textbf{Evaluation of RCPP.} We consider an optimization and a calibration dataset with samples of $w^{(1)}, \hdots, w^{(K + L)}$ from a normal distribution, $P_Y \coloneq \mathcal{N}(0, 0.01^2) \times T$. The test data $w$ is drawn from the distribution $P_{\tilde{Y}} \coloneq \mathcal{N}(0, 0.013^2) \times T$, which simulates a distribution shift from $P_Y$. We select KL divergence as the choice of $f$-divergence where $\phi(t) \coloneq tlog(t)$. We follow \cite{devroye2018total} and compute $\epsilon \coloneq 0.079$ via $D_\phi(\mathcal{N}(\mu_1, \Sigma_1), \mathcal{N}(\mu_2, \Sigma_2)) = \epsilon \coloneq \frac{1}{2}[\text{tr}(\Sigma_1^{-1}\Sigma_2-I) + (\mu_1 - \mu_2)^T\Sigma_1^{-1}(\mu_1 - \mu_2) - \text{logdet}(\Sigma_2 \Sigma_1^{-1})]$ given that $\Sigma_1$ and $\Sigma_2$ are positive definite and denote the covariance matrices of $P_{\tilde{Y}}$ and $P_Y$ respectively and $\mu_1 \coloneq \mu_2 \coloneq 0 \in \mathbb{R}^{T \times 4}$.

We fix the parameters $N \coloneq 100, V \coloneq 1000, K \coloneq 60$ and $L \coloneq 200$ and conduct the evaluation on the marginal feasibility guarantee with CPP-MIP. We compute $EC$ with $\tilde{C}(x^*_l)$ (where $x^*_l$ is solved with $\alpha_r$) from Theorem \ref{thm:rcpp_marginal} and with $C_m(x^*_l)$ (where $x^*_l$ is solved with $\alpha_m$) from Theorem \ref{thm:a_posteriori_marginal} as a baseline. We observe an $EC$ of 0.80 with $C_m$ and 0.96 with $\tilde{C}$, where the baseline undercovers in comparison to  RCPP. We also show in Figure \ref{fig:case3} (a) the histogram of $\tilde{C}(x_l^*)$ and $C_m(x_l^*)$ where $C_m(x_l^*)$ are in general less than $\tilde{C}(x_l^*)$ as expected. We show in Figure \ref{fig:case3} (b) the attained optimal costs from CPP-MIP with RCPP and the baseline methods.

\textbf{Evaluation of Mondrian CPP.} We fix the parameters  $N \coloneq 200, V \coloneq 1000, K \coloneq 60$ and $L \coloneq 200$ and conduct the evaluation on the marginal feasibility guarantee with CPP-MIP. We consider again a distribution $P_Y \coloneq \mathcal{N}(0, 0.01^2) \times T$. We divide the support of $P_Y$ into two groups representing disturbances of small and large magnitudes respectively. Specifically, let $G_1 \coloneq [-0.027, 0.027] \times T$, $G_2 \coloneq (-\infty, \infty) \times T \setminus G_1$, and $\mathcal{G} \coloneq \{G_1, G_2\}$. We observe that $MEC(C_m, G_2) = 0.86$ and $MEC(C_{G_2}, G_2) = 0.94$ where $MEC(C_m, G_2)$ undercovers as compared to $MEC(C_{G_2}, G_2)$. We show the histogram of $C_m(x_l^*)$ and $C_G(x_l^*)$ in Figure \ref{fig:case3} (c).

\section{Conclusion}\label{sec:conclusion}
We proposed a new framework, conformal predictive programming (CPP), for chance constrained optimization (CCO) problems. CPP is built on conformal prediction, an uncertainty quantification tool. We showed how to obtain marginal and conditional feasibility guarantees of the CPP solution for the CCO problem. We argued that CPP can easily incorporate other variants of CCO, which we illustrated using robust and Mondrian CP. 

\bibliographystyle{plain}        
\bibliography{autosam}          


\section{Appendix}
\label{sec:appendix}

\subsection{Robust Conformal Prediction}
  Recall that $R, R^{(1)}, \hdots, R^{(L)}$ was assumed in Assumption \ref{assumption:cp} to be identically distributed. In practice, however, this assumption may be violated, e.g., we may  have calibration data $R^{(1)}, \hdots, R^{(L)}$ from a simulator while the data $R$ observed during deployment is different. Nonetheless, we would like to provide guarantees when $R$ and $R^{(1)}, \hdots, R^{(L)}$ are statistically close. Let $P_R$ and $P_{R_0}$ denote  calibration and deployment distributions, respectively, and let $ R^{(1)}, \hdots, R^{(L)} \sim P_R$ while $R\sim P_{R_0}$. To capture their distance, we use the $f$-divergence, $D_\phi(P_{R_0},P_R)\coloneqq \int_\mathcal{X} \phi\Big(\frac{d P_{R_0}}{d P_R}\Big) d P_R$, where $\mathcal{X}$ is the support of $P_R$ and where $\frac{d P_{R_0}}{d P_R}$ is the Radon-Nikodym derivative. It is hence assumed that $P_{R_0}$ is  absolutely continuous with respect to $P_R$. The function $\phi: [0, \infty) \rightarrow \mathbb{R}$ needs to be convex with $\phi(1) = 0$ and $\phi(t) < \infty$ for all $t > 0$. If $\phi(z) \coloneqq \frac{1}{2}|z - 1|$, we attain the total variation distance $TV(P_{R_0}, P_R) \coloneqq \frac{1}{2}\int_x|P(x) - Q(x)|dx$ where $P$ and $Q$ represent the probability density functions corresponding to $P_{R_0}$ and $P_R$. The next result is mainly taken from \cite{cauchois2024robust} and is presented as summarized in \cite{zhao2024robust}. \\
\begin{lemma} \textbf{Robust Quantile Lemma.} [Corollary 2.2 in \cite{cauchois2024robust}] \label{lemma:robust_cp}
    Let $R^{(1)},\hdots,R^{(L)}\sim P_R$ and $R \sim P_{R_0}$ be independent random variables such that $D_\phi(P_{R_0},P_R)\le \epsilon$. For a failure probability of $\delta\in (0,1)$, assume that $L \ge \Big\lceil \frac{v^{-1}(1 - \delta)}{1 - v^{-1}(1 - \delta)} \Big\rceil$ with
    \begin{align*}
        \alpha_r(L) & \!\coloneqq\! v^{-1}(1-\delta_n(L)), \\
        \delta_n(L) & \!\coloneqq\! 1-v\big((1+1/L)v^{-1}(1-\delta)\big),\\
        v(\beta) & \!\coloneqq\! \inf \{z\!\in\![0,1] \!\mid \! \beta \phi(z/\beta)\!+\! (1\!-\!\beta)\phi(\frac{1\!-\!z}{1\!-\!\beta}) \!\le\! \epsilon\},\\
        v^{-1}(\tau) & \!\coloneqq\! \sup \{\beta\in[0,1] \mid v(\beta)\le \tau\}.
    \end{align*}
    Then,  it holds that $\tilde{\mathbb{P}}_m(R \le \tilde{C})\ge 1-\delta$ with
    \begin{align}\label{eq:C_tilde}
        \tilde{C} \coloneqq \hat{Q}_{\alpha_r(L)}(R^{(1)},\hdots,R^{(L)}).
    \end{align}
\end{lemma}
\vspace{-8pt}
We emphasize that computation of $v$ and $v^{-1}$ in Lemma \ref{lemma:robust_cp} is efficient as it involves solving convex optimization problems. A similar result was presented in \cite{aolaritei2025conformal}, but using the  L\'evy-Prokhorov metric instead of an $f$-divergence. We could also use this variant for our robust CCP version, illustrating again the versatility of our framework.

\subsection{Feasibility and Optimality Guarantees of Sample Average Approximation}
\label{app:other_guarantees}
We here illustrate the a priori feasibility and optimality guarantees from SAA, which apply to our proposed algorithm of CPP. Let us denote the feasibility region of the SAA problem in \eqref{eq:saa} as $F_\omega(K) \coloneq \{x \mid x \in \mathcal{X} \text{ and } x \models \eqref{eq:saa}\}$ and the optimal solution as $x^*_\omega(K)$. SAA produces a priori conditional feasibility guarantees for $F_\omega \subseteq F$ and optimality guarantees for $J(x^*_\omega) \le J(x^*)$. \\
\begin{theorem}\label{thm:optimality}
    \textbf{a priori Optimality Guarantee [Lemma 1 from \cite{luedtke2008sample}]} Let Assumption \ref{assumption:cp} hold. Suppose $\eqref{eq:cco}$ has an optimal solution $x^*$. For any SAA optimal solution $x^*_\omega$, it holds that
    \begin{align*}
        \mathbb{P}(J(x^*_\omega) \le J(x^*)) \ge \sum_{i = 0}^{\lfloor \omega K \rfloor}\binom{K}{i}\delta^i(1 - \delta)^{K - i}.
    \end{align*}
\end{theorem}
In fact, when $K$ grows, Theorem \ref{thm:optimality} is only desirable if $\omega > \delta$ as shown in \cite{luedtke2008sample}. Since this does not apply to CPP (where in Remark \ref{remark:connection_saa} we have $\omega < \delta$), we remark that readers interested more in optimality than feasibility should choose a different $\omega$ than the one proposed in this work, which focuses on feasibility guarantees.\\

\begin{theorem}
    \textbf{a priori Feasibility Guarantees [Theorems 5, 8, 9 and 10 from \cite{luedtke2008sample}]}\label{thm:a_priori_feasibility} Let Assumption \ref{assumption:cp} hold. Suppose $\omega \in [0, \delta)$, then the following holds
    \begin{itemize}
        \item If $\mathcal{X}$ is finite, then we have
        \begin{align*}
            \mathbb{P}_K(F_\omega \subseteq F) \ge 1 - |\mathcal{X} \setminus F|\exp(-2K(\delta - \omega)^2)
        \end{align*}
        where $\mathcal{X} \setminus F$ denotes set substraction.
        \item If $f(x, Y) \coloneq Y - g(x)$ for some $g: \mathbb{R}^n \rightarrow \mathbb{R}^d$ and $Y$ has a finite distribution (i.e. $Y$ has a support of $\Xi = \{Y^1, \hdots, Y^H\}$ for $H \in \mathbb{N}$), then we have
        \begin{align*}
            \mathbb{P}_K(F_\omega \subseteq F) \ge 1 - |\prod_{j = 1}^d\Xi_j|\exp(-2K(\delta - \omega)^2)
        \end{align*}
        where $\Xi_j \coloneq \{Y_j^h: h = 1, \hdots, H\}$ where $Y^h_j$ denotes the $j$-th compoinent of $Y^h$.
        \item If $f(x, Y) \coloneq Y - g(x)$ for some $g: \mathbb{R}^n \rightarrow \mathbb{R}^d$ and $F \subseteq \overline{X}(l, u) \coloneq \{x \in \mathcal{X} \mid l \le g(x) \le u\}$ for some $l, u \in \mathbb{R}^d$ and $g$ is $\mathcal{L}$-Lipschitz, it holds that 
        \begin{align*}
            \mathbb{P}_K(F_\omega(l, u) \subseteq F) \ge& 1 - \lceil D\mathcal{L} /\kappa \rceil^d\exp(-2K(\delta - \\ &\omega - \kappa)^2)
        \end{align*}
       for any $\kappa \in (0, \delta - \omega)$ and $D \coloneq \max \{u_j - l_j, j = 1, \hdots, d\}$ where $F_\omega(l, u) \coloneq \{x \in \overline{X}(l, u) \text{ s.t. } x \models \eqref{eq:saa}\}$.
        \item Let $\mathcal{X}$ be bounded with diameter $D \coloneq \sup\{\|x - x'\|_\infty: x, x' \in \mathcal{X}\}$ and $f$ is $\mathcal{L}$-Lipschitz. For any $\kappa \in (0, \delta - \omega)$ and $\theta  > 0$,
        \begin{align*}
            \mathbb{P}_K(F_{\omega, \theta} \subseteq F) &\ge 1 - \lceil\frac{1}{\kappa}\rceil\lceil 2 \mathcal{L}D/\theta\rceil^n\exp(-2K(\delta \\&- \omega - \kappa)^2),
        \end{align*}
        where $F_{\omega, \theta} \coloneq \{x \in \mathcal{X} \mid \frac{1}{K}\sum_{i = 1}^K\mathbbm{1}\{f(x, Y^{(i)}) + \theta \le 0\} \ge 1 - \omega\}$.
    \end{itemize}
\end{theorem}
We emphasize that optimizing with the slack variable $\theta$ or a restricted domain $F_\omega(l, u)$ via the substitution in Remark \ref{remark:connection_saa} does not hinder the validity of our a posteriori feasibility guarantees, which will be made more efficient however via choosing a small $\theta$ or if $F_\omega$ is tight on $F_\omega(l, u)$ for reasons elaborated in Remark \ref{remark:marginal}. We remark that results in Theorem \ref{thm:optimality} and \ref{thm:a_priori_feasibility} also apply to joint chance constraints \cite{luedtke2008sample}.

\subsection{Computational Encoding of the Quantile}
\label{sec:encoding}
We present three approaches through which the quantile in equation \eqref{eq:cp_quantile_cons} can be computed efficiently. We first present a mixed-integer programming approach (MIP) adapted from \cite{geng2019data}. The MIP approach (which we refer to as CPP-MIP) reformulates the quantile within the optimization problem \eqref{eq:cp_quantile} with a set of mixed integer constraints. A feasible solution to CPP-MIP is also a feasible solution to \eqref{eq:cp_quantile}, and vice versa. However, the necessity of integer variable makes the problem NP-hard. Motivated by this observation, we further propose CPP-Bilevel. CPP-Bilevel is based on representing the quantile within the optimization problem \eqref{eq:cp_quantile} as a linear optimization problem, which leads to a bilevel optimization problem which we then solve by reformulating the inner program with its KKT conditions. A feasible solution to CPP-Bilevel is also a feasible solution to \eqref{eq:cp_quantile}, while the other direction only holds under some assumptions.  Lastly, we propose another reformulation, inspired by \cite{campi2011sampling}, in the convex setting that accurately captures the quantile by discarding restrictive constraints (which we refer to as CPP-Discarding).
For simplicity, we set $\alpha=\alpha_m$ in this settings, but we remark that all results apply without loss of generality to quantile reformulations with a general quantile level of $\alpha \in (0, 1)$, and thus other  conformal prediction variants.

\subsubsection{Quantile Encoding with Mixed Integer Programming.} \label{subsec:mip}
We first summarize the rewriting of the quantile in equation \eqref{eq:cp_quantile} using mixed integer programming (MIP), adapted from \cite{geng2019data}.

We start by introducing the MIP encoding from \cite{bemporad1999control}. Consider a real-valued function $\mu(x)$ and a binary variable $z \in \{0, 1\}$. Then, the mixed integer linear constraint\st{s}
\begin{subequations} \label{eq:encoding_MILP}
    \begin{align}
        \mu(x) \le M(1 - z),
    \end{align}
    \end{subequations}
    enforces that $\mu(x) \le 0$ if  $z = 1$  where $M \in \mathbb{R}$ is a sufficiently large positive constant, see \cite{bemporad1999control} for  details.

Following the same reasoning as equation \eqref{eq:cp_quantile_cons}, we recall that the quantile constraint in \eqref{eq:cp_quantile_cons} is equivalent to
\begin{align}\label{eq:quantile2mip}
    \sum_{i = 1}^K \mathbbm{1} \{f(x, Y^{(i)}) \le 0\} \ge \lceil(K + 1)(1-\delta)\rceil=\lceil K\alpha \rceil.
\end{align}
We proceed by introducing binary variables $z_i \in \{0,1\}$ for $ i\in \{1, \dots, K\}$ that encode the satisfaction of $f(x, Y^{(i)}) \le 0$ along with a set of mixed integer constraints. Concretely, we present CPP-MIP as
\begin{subequations}\label{eq:cp_mip}
\begin{align}
    & \min_{x \in \mathcal{X}, z \in \{0, 1\}^K} \quad J(x)\\
    \textrm{s.t.} & \quad f(x, Y^{(i)})  \leq M (1 - z_i), i \in \{1, \dots, K\},   \label{cons_mip_2}\\
    & \quad \sum_{i=1}^K z_i \geq \lceil K\alpha \rceil, \label{cons_mip_4}
\end{align}
\end{subequations}
where $M = \max_{i = 1, \hdots, K}\max_{x \in \mathcal{X}} f(x, y^{(i)})$. We note that an over-approximation of $M$ suffices, see \cite{bemporad1999control}, and that $M$ exists when $\mathcal{X}$ is a compact set and $f$ is continuous\footnote{In our case studies, we also enforce $f(x, Y^{(i)}) \geq \zeta + (m - \zeta) z_i, i \in \{1, \dots, K\}$, where $\zeta \in \mathbb{R}$ is a small positive constant, e.g., machine precision, and $m = \min_{i = 1, \hdots K}\min_{x\in \mathcal{X}}f(x, y^{(i)})$ on top of \eqref{cons_mip_2} and \eqref{cons_mip_4} in \eqref{eq:cp_mip}, because such a design choice preserves the feasibility implication in Theorem \ref{thm:mip} (any point feasible for the augmented formulation remains feasible for \eqref{eq:cp_mip}) and at the same time forces the binary variables $z_i$ to exactly encode the nonpositivity of $f(x, Y^{(i)})$, allowing more transparent numerical verification and semantic interpretations. Under-approximation of $m$ suffices.}. The next result establishes the equivalence between the optimization problems in equations \eqref{eq:cp_quantile} and  \eqref{eq:cp_mip}. In this paper, we say that two programs are equivalent if they share the same optimal solution $x^*$. It follows immediately from the previous construction and is provided without a proof. \\

\begin{theorem}\label{thm:mip}
    The optimization problem in \eqref{eq:cp_mip} is equivalent to the optimization problem \eqref{eq:cp_quantile}.
\end{theorem}

We emphasize that solving MIP problems, such as in \eqref{eq:cp_mip}, are in general NP-hard. However, these problems can usually be solved efficiently in practice, e.g., using optimization solvers such as SCIP \cite{achterberg2009scip},  rarely encountering the worst case complexity, as we demonstrate in Section \ref{sec:case_studies}. Note also that the optimization problem in \eqref{eq:cp_mip} reduces to a mixed integer linear program when $J$ and $f$ are affine in $x$ for all $Y$ and when $\mathcal{X}$ is parameterized by affine functions. Nevertheless, given that CPP-MIP is in general difficult to solve theoretically, we are motivated to present CPP-Bilevel as an alternative.

\subsubsection{Quantile Encoding with Bilevel Optimization} Following ideas from \cite{koenker1978regression, cleaveland2024conformal},  we now rewrite the quantile constraint in equation \eqref{eq:cp_quantile_cons} as the linear program 
\begin{subequations}\label{eq:approx_}
\begin{align}
 & q^* = \argmin_{q} \sum_{i=1}^K(\alpha e^+_i + (1 - \alpha)e^-_i) \label{eq:3e_}\\
    \quad & \textrm{s.t. } e^+_i - e^-_i = f(x, Y^{(i)}) - q, \label{eq:3f_}\\
    \quad & \quad \ \ \,  e^-_i, e^+_i \ge 0, \forall i \in \{1, \hdots, K\}, \label{eq:3g_}
\end{align}
\end{subequations}
where $q, e_i^+, e_i^-\in\mathbb{R}$ are decision variables. Intuitively, the optimization problem in \eqref{eq:approx_} minimizes a weighted sum of the distance between the $\alpha$-quantile $q$ and each sample $f(x, Y^{(1)}),\hdots,f(x, Y^{(K)})$. We show how the solution $q^*$ of \eqref{eq:approx_} relates to the quantile constraint \eqref{eq:cp_quantile_cons}.\\
\begin{lemma}\label{lem:over}
    It holds that  $\hat{Q}_{\alpha}(f(x, Y^{(1)}), \hdots,f(x, Y^{(K)}))$ $\le q^*$, i.e., the solution $q^*$ to \eqref{eq:approx_} upper bounds the  quantile constraint \eqref{eq:cp_quantile_cons}. Equivalence holds if $\alpha K \notin \mathbb{N}$.
    \begin{proof}
    Consider the function $\rho_\alpha(u) := u(\alpha - \mathbbm{1}(u < 0))$ and the optimization problem
    \begin{equation}
    \label{eq:check_function}
    \begin{aligned}
        \argmin_q \sum_{i = 1}^K \rho_\alpha(f(x, Y^{(i)}) - q).
    \end{aligned}        
    \end{equation}
    
    Note that $\rho_\alpha(u)$ is convex in $u$, and thus $\rho_\alpha(f(x,Y^{(i)}) - q)$ is convex in $q$. Therefore, the objective in \eqref{eq:check_function} is convex in $q$, and the subgradient optimality condition is necessary and sufficient for optimality.
    Let $F(z) = \frac{1}{K} \sum_{i = 1}^K \mathbbm{1}(f(x, Y^{(i)}) \le z)$ denote the empirical cumulative distribution function over $f(x, Y^{(1)}), \hdots,\\ f(x, Y^{(K)})$. By the subgradient optimality condition, we know that the solution $q^*$ of \eqref{eq:check_function} satisfies $0 \in \partial_q\sum_{i = 1}^K \rho_\alpha(f(x, Y^{(i)}) - q^*)$, where we denote by $\partial_q$ the subdifferential with respect to $q$.
    
    Let $N_{q^*}^\sim \coloneq \sum_{i = 1}^K\mathbbm{1}(f(x, Y^{(i)}) \sim q^*)$ where $\sim \in \{<, >, =\}$. Then,
    \begin{align*}
    0 &\in \{(1 - \alpha)N_{q^*}^< - \alpha N_{q^*}^> \} \oplus [-\alpha, 1 - \alpha]N_{q^*}^= \\
      &= [N_{q^*}^< - \alpha K,\; N_{q^*}^< + N_{q^*}^= - \alpha K],
\end{align*} \noindent where $\oplus$ denotes the Minkowski sum.
    
    Equivalently, $\sum_{i = 1}^K\mathbbm{1}(f(x, Y^{(i)}) \le q^*) \ge \alpha K \ge \sum_{i = 1}^K\mathbbm{1}(f(x, Y^{(i)}) < q^*)$.  
    
    It is easy to see, and pointed out in \cite[Chapter 1]{koenker2005quantile}, that if $\alpha K \notin \mathbb{N}$, we have a unique minimizer for \eqref{eq:check_function} at $q^* = \hat{Q}_{\alpha}(f(x, Y^{(1)}), \hdots, f(x, Y^{(K)}))$. If $\alpha K \in \mathbb{N}$, the minimizer may be non-unique. However, the optimality condition above implies that every minimizer $q^*$ satisfies $F(q^*)\ge \alpha$. Since $\hat{Q}_{\alpha}(f(x,Y^{(1)}),\hdots,f(x,Y^{(K)}))=\inf\{z\in\mathbb{R}:F(z)\ge \alpha\}$, it follows that $\hat{Q}_{\alpha}(f(x,Y^{(1)}),\hdots,f(x,Y^{(K)}))\le q^*$.
    
    Finally, we need to show that \eqref{eq:approx_} is equivalent to \eqref{eq:check_function}. Note that 
    $\argmin_q \sum_{i = 1}^K \rho_\alpha(f(x, Y^{(i)}) - q) 
    = \argmin_q \big(\sum_{i=1}^K\alpha (f(x, Y^{(i)}) - q)\mathbbm{1}(f(x, Y^{(i)}) \ge q) 
    + \sum_{i = 1}^K(\alpha - 1)(f(x, Y^{(i)}) - q)\mathbbm{1}(f(x, Y^{(i)}) < q)\big)$, 
    which is equivalent to \eqref{eq:approx_} by variable splitting.
\end{proof}
\end{lemma}\raggedbottom

We can now use the linear program in \eqref{eq:approx_} to replace equation \eqref{eq:cp_quantile_cons}, resulting in CPP-Bilevel
\begin{subequations}\label{eq:cp_bilevel}
\begin{align}
    \min_{x \in \mathcal{X}} \quad & J(x)\\
    \textrm{s.t.} \quad & q^* \le 0,\\
                        \quad & \eqref{eq:3e_}, \eqref{eq:3f_},\eqref{eq:3g_}.
\end{align}
\end{subequations}
Denote the feasibility region of \eqref{eq:cp_bilevel} by $F_{b}(K)\subseteq \mathbb{R}^n$. Using Lemma \ref{lem:over}, we obtain the following result.\\
\begin{corollary}\label{thm:quantile_linear}
    For any choice of $K \in \mathbb{N}$, $F_{b}(K) \subseteq F(K)$. If $\alpha K \notin \mathbb{N}$, $F_{b}(K) = F(K)$. 
\end{corollary}
Note that the inner optimization problem in equation \eqref{eq:cp_bilevel} is composed of equations \eqref{eq:3e_}, \eqref{eq:3f_}, and \eqref{eq:3g_}.  For any fixed value of the decision variable $x$ from the outer optimization problem, the inner optimization problem is linear in $q, e^+, \text{ and }e^-$. We can hence rewrite the inner optimization problem with its KKT conditions \cite{boyd2004convex}. This results in the optimization problem 
\begin{subequations}\label{eq:cp_kkt}
    \begin{align}
    & \min_{x \in \mathcal{X}, \gamma, \lambda, \beta, q, e^-, e^+} \quad  J(x) \label{eq:4a}\\
    \textrm{s.t. } &\qquad q \le 0, \label{eq:4b}\\
        & \quad \alpha + \gamma_i - \lambda_i = 0, i \in \{1, \hdots, K\},\label{eq:4e}\\
        & \quad 1 - \alpha - \gamma_i -\beta_i = 0, i \in \{1, \hdots, K\},\label{eq:4f}\\
        & \quad \sum_{i = 1}^K \gamma_i = 0,\label{eq:4g}\\
        & \quad e^+_i \!-\! e^-_i \!-\! f(x, Y^{(i)}) \!+\! q \!=\! 0, i \!\in\! \{1, \hdots, K\},\label{eq:4h}\\
        & \quad e^-_i, e^+_i \ge 0, i \in \{1, \hdots, K\},\label{eq:4i}\\
        & \quad \lambda_i, \beta_i \ge 0, i \in \{1, \hdots, K\},\label{eq:4j}\\
        & \quad  \lambda_ie_i^+ = 0, i \in \{1, \hdots, K\},\label{eq:4k}\\
        & \quad \beta_ie_i^- = 0, i \in \{1, \hdots, K\},\label{eq:4l}
    \end{align}
    \end{subequations}
    where $\beta_i, \gamma_i, \lambda_i \in \mathbb{R}$ are new decision variables.
    Specifically, \eqref{eq:4b} denotes the quantile constraint from the outer optimization problem, while \eqref{eq:4e}-\eqref{eq:4g} represent the stationarity condition, \eqref{eq:4h}-\eqref{eq:4i} denote primal feasibility conditions, \eqref{eq:4j} denotes dual feasibility condition, and \eqref{eq:4k}-\eqref{eq:4l} denote complementary slackness condition. We summarize our main result next.\\

\begin{theorem}\label{thm:linear_kkt}
    The optimization problem in \eqref{eq:cp_kkt} is equivalent to \eqref{eq:cp_bilevel}. A feasible solution to \eqref{eq:cp_kkt}, excluding the auxiliary variables (variables other than $x$), is a feasible solution to \eqref{eq:cp_quantile} and the reverse holds if $\alpha K \notin \mathbb{N}$.
    \begin{proof}
        A linear program has zero duality gap \cite{boyd2004convex}. This implies that the optimal solution of the inner problem in equations \eqref{eq:3e_}, \eqref{eq:3f_}, and \eqref{eq:3g_} is equivalent to  the KKT conditions in \eqref{eq:4e}-\eqref{eq:4l}. Hence, \eqref{eq:cp_kkt} is equivalent to \eqref{eq:cp_bilevel}. The rest applies from Corollary \ref{thm:quantile_linear}.
    \end{proof}
\end{theorem}
Note that \eqref{eq:cp_kkt} is a nonconvex optimization problem even when $J$, $h_i$, $g_i$ are convex and $f$ is an affine function  due to  constraints \eqref{eq:4k} and \eqref{eq:4l}. However, in this case \eqref{eq:cp_kkt} is a linear complimentarity program for which efficient solvers exist \cite{fischer1995newton}. We remark that local optima of \eqref{eq:cp_kkt} do not generally correspond to  local optima of \eqref{eq:cp_bilevel}, see \cite{kleinert2021survey}, and that feasible solutions to \eqref{eq:cp_kkt} violate standard constraint qualifications \cite{ye1995optimality}. Therefore, heuristic algorithms such as branch-and-cut solutions are developed for tractable solutions \cite{kleinert2021survey}.

\subsubsection{Quantile Encoding with Discarding}

\begin{algorithm}[t]
\caption{CPP-Discarding for Quantile Encoding}
\label{alg:cpp-discarding}
\begin{algorithmic}[1]
\Require Objective $J(x)$, constraints $f(x,Y^{(i)}) \le 0$ for $i=1,\dots,K$, quantile level $\alpha \in (0,1]$
\Ensure Feasible solution $x^\star$

\State Initialize index set $\mathcal{I} \gets \{1,\dots,K\}$
\While{true}
    \State Solve
    \[
        x^\star \in \arg\min_{x \in \mathcal{X}} J(x)
        \quad \text{s.t. } f(x,Y^{(i)}) \le 0,\ \forall i \in \mathcal{I}
    \]
    \State Find active constraints at $x^\star$:
    \[
        A \gets \{ i \in \mathcal{I} \mid f(x^\star,Y^{(i)}) = 0 \}
    \]
    \If{$A = \emptyset$}
        \State \textbf{break}  \Comment{all remaining constraints are inactive}
    \EndIf
    \State Choose one active constraint $i' \in A$
    \State Update the index set:
    \[
        \mathcal{I} \gets \mathcal{I} \setminus \{i'\}
    \]
    \If{$|\mathcal{I}| = \lceil K\alpha \rceil$}
        \State \textbf{break}
    \EndIf
\EndWhile
\State \Return $x^\star$
\end{algorithmic}
\end{algorithm}

Inspired by sampling-and-discarding SA from \cite{campi2011sampling}, we propose a method to solve the quantile reformulation by iteratively solving a series of convex programs. As discussed in Section \ref{subsec:mip}, the quantile constraint in \eqref{eq:cp_quantile_cons} requires that at least $\lceil K \alpha \rceil$ of the $K$ constraints $f(x, Y^{(i)}) \le 0$ are satisfied, as formulated in equation \eqref{eq:quantile2mip}.
To achieve this, we iteratively solve the following convex optimization problem:
\begin{subequations}\label{eq:cp_discarding}
\begin{align}
    \min_{x \in \mathcal{X}} \quad & J(x)\\
    \textrm{s.t.} \quad & f(x, Y^{(i)}) \le 0, \forall i \in \mathcal{I},
\end{align}
\end{subequations}
where $\mathcal{I}$ is the index set of each iteration. Initially, we set $\mathcal{I} = \{1, \dots, K\}$ to include all constraints. 
After solving this optimization problem, we identify one active constraint, i.e., one for which $f(x, Y^{(i')}) = 0$ for some $i'\in \mathcal{I}$, and remove $i'$ from the set $\mathcal{I}$. We repeat this process until one of the following stopping conditions is met: (1) all remaining constraints are inactive,  or (2) $|\mathcal{I}| = \lceil K \alpha \rceil$. 
We summarize the process in Algorithm~\ref{alg:cpp-discarding}.
CPP-Discarding is sound, as shown next.\\ 

\begin{remark}\label{thm:discarding}
    If the optimization problem \eqref{eq:cp_discarding} is initially feasible for $\mathcal{I} = \{1, \dots, K\}$, then the optimal solution to CPP-Discarding (i.e., the aforementioned discarding framework) satisfies the constraint \eqref{eq:cp_quantile_cons}.
    \begin{proof}
    The optimal solution obtained through CPP-Discarding ensures that at least $\lceil K \alpha \rceil$ out of the $K$ constraints $f(x, Y^{(i)}) \le 0$ are satisfied, guaranteeing that the quantile constraint \eqref{eq:cp_quantile_cons} is satisfied.
    \end{proof}
\end{remark}

This soundness result is trivial. In fact, one can discard any constraint (not necessarily the active ones) and Remark \ref{thm:discarding} will still hold. Our choice of removing active constraints is motivated in the convex setting.\\

\begin{assumption}
\label{assumption:convexity_discard}
    Assume that the constraint and cost functions $f(x, Y)$ and $J(x)$ are convex in the argument $x$  and that $\mathcal{X}$ is a convex set.
\end{assumption}

Under Assumption \ref{assumption:convexity_discard}, we note that in CPP-Discarding we either terminate when (1) all remaining constraints are inactive in which case the global optimal value of $J(x)$ has been achieved, or (2) $|\mathcal{I}| = \lceil K \alpha \rceil$ in which case $K - \lceil K \alpha \rceil$ constraints have been removed. Our strategy guarantees that each removal generates a candidate solution that is no worse than the previous iteration.


Note that, unlike Theorems \ref{thm:mip} and \ref{thm:linear_kkt} where, under some conditions, CPP-MIP and CPP-Bilevel are equivalent to the CPP problem \eqref{eq:cp_quantile}, we can here only guarantee that the obtained optimal solution to CPP-Discarding is a feasible solution to the CPP problem \eqref{eq:cp_quantile}. The optimal solution to CPP-Discarding is not optimal to \eqref{eq:cp_quantile} unless an early termination occurs. The reason lies in the possibility of multiple active constraints: discarding different constraints can lead to different solutions, thereby losing the guarantee of achieving global optimality. Furthermore, we note that the optimization problem \eqref{eq:cp_discarding} can initially be infeasible since we require that all $K$  constraints are satisfied simultaneously.

Finally, we conclude this section by comparing the three encodings. 
In the convex setting, CPP-Discarding has the lowest complexity due to its convexity. However, it is prone to infeasibility at the initial stage, which can render the framework inapplicable. As for CPP-Bilevel and CPP-MIP, a theoretical comparison of their performance is challenging and we leave such comparison as future work.

\subsection{Proofs and Auxiliary Statements}

In this section, we illustrate all proofs for the theorems/lemmas and auxiliary results in this paper.

\subsubsection{Proofs for Theorems and Lemmas}

 \begin{proof}
 \textbf{Proof for Lemma \ref{lemma:separable}.} We only provide the proof for the case $\alpha(K)=\alpha_m(K)$, while the proof for the case $\alpha(K)=\alpha_c(K)$ follows similarly. For the specific choice of the function $f$, note that the constraint \eqref{eq:cp_quantile_cons} is equivalent to $\hat{Q}_{\alpha_m(K)}(h(Y^{(1)}), \hdots,h(Y^{(K)})) \!\le\! g(x)$. By Lemma \ref{lemma:quantile_lemma}, $\mathbb{P}^{K + 1}(h(Y) \le \hat{Q}_{\alpha_m(K)}(h(Y^{(1)}), \hdots,h(Y^{(K)}))) \\\ge 1 - \delta$, since $h(Y^{(1)}), \hdots, h(Y^{(K)})$ are independent.
 \end{proof}

\begin{proof}
\textbf{Proof for Theorem \ref{thm:a_posteriori_marginal}.} The solution $x^*(K)$ trivially satisfies $x^*(K) \in \mathcal{X}$. Since $x^*(K)$ is independent from $Y^{(K + 1)},\hdots,Y^{(K + L)}$ and since $Y^{(K + 1)},\hdots,Y^{(K + L)}$ are i.i.d. by Assumption \ref{assumption:cp_post}, it also follows that $f(x^*, Y^{(K + 1)}), \hdots, f(x^*, Y^{(K + L)})$ are i.i.d. Then, by Lemma \ref{lemma:quantile_lemma}, we can conclude that $\mathbb{P}_m(f(x^*(K), Y) \le C_m(x^*(K))) \ge 1 - \delta$. 
    \end{proof}

\begin{proof} 
\textbf{Proof for Theorem \ref{thm:quantile_shift}.} Note that Lemma \ref{lemma:cond_quantile_lemma} guarantees that 
    \begin{align*}
        \mathbb{P}_L(\mathbb{P}(f(x^*{(K)}, Y) \le C_c(x^*(K))) \ge 1 - \delta^*) \ge 1 - \beta.
    \end{align*}
    Now, note that ensuring $C_c(x^*(K)) \le 0$ is equivalent to
    \begin{align*}
        S \ge \lceil(L + 1)(1-\delta+\sqrt{\frac{\ln{(1/\beta)}}{2L}})\rceil.
    \end{align*}
        We hence observe that  $\delta^* \coloneq \min\{\delta' \mid  S\ge  \lceil (L + 1)(1 - \delta' + \sqrt{\frac{\ln(1/\beta)}{2L}})\rceil \}$ corresponds to the minimum probability that ensures $C_c(x^*(K)) \le 0$. From here, we obtain $\delta^* = 1 - \frac{S}{L + 1} + \sqrt{\frac{\ln(1/\beta)}{2L}}$ by simple manipulation.
    \end{proof}

\begin{proof}
\textbf{Proof for Theorem \ref{thm:rcpp_marginal}.} As in the proof of Theorem \ref{thm:a_posteriori_marginal}, we note that $f(x^*(K), Y^{(K + 1)}), \hdots, \\f(x^*(K), Y^{(K + L)}) \sim P_R$ are i.i.d, where $P_R$ is the pushforward distribution of $P_Y$ under $f(x^*(K),\cdot)$. Let now $P_{Y'}\in \mathcal{P}(P_Y, \epsilon)$ and  $Y \sim P_{Y'}$. Further, let $P_{R_0}$ be the pushforward distribution of $P_{Y'}$ under $f(x^*(K),\cdot)$, i.e., $f(x^*(K), Y)\sim P_{R_0}$. By the data processing inequality, it follows that $D_\phi(P_{R_0}, P_R) \le \epsilon$. Therefore, by Lemma \ref{lemma:robust_cp}, we have $\tilde{\mathbb{P}}_m(f(x^*(K), Y) \le \tilde{C}(x^*(K))) \ge 1 - \delta$.
\end{proof} 

\subsubsection{Auxiliary Statements}
 We note that $\mathbb{P}(R \le C_m)$ is by itself a random variable, which is discussed in \cite{angelopoulos2021gentle, lindemann2025formal} without a proof. We summarize it below. Similarly $C_m$ is also a random variable dependent on the calibration set.\\
 
\begin{lemma}\label{lemma:marginal_2_cond}
If $P_R$ is a continuous distribution (i.e. if the nonconformity scores are distinct almost surely), $\mathbb{P}(R \le C_m) \sim \text{Beta}(L + 1 - l, l)$ with $l \coloneq \lfloor(L + 1)\delta\rfloor$ where $\text{Beta}(\cdot)$ denotes the Beta distribution. 
\begin{proof} From the proof of \cite[Proposition 2a]{vovk2012conditional}, we know that $\mathbb{P}_L(\mathbb{P}(R > C_m) > \delta) \le \mathbb{P}_L(B \le \lfloor \delta(L + 1) - 1 \rfloor)$ where $B$ is a binomial random variable with parameters $L, \delta$.\footnote{Here, L denotes the total number of Bernoulli trials and $\delta$ denotes the success probability of each Bernoulli experiment.} Here, equality holds if $P_R$ is continuous. Under the continuity assumption, we have $\mathbb{P}_L(\mathbb{P}(R \le C_m) \ge 1 - \delta) = 1 - bin_{L, \delta}(\lfloor \delta(L + 1) - 1 \rfloor)$ where $bin_{L, \delta}$ is the cumulative binomial distribution function. Then,
\begin{align*}
    & bin_{L, \delta}(\lfloor \delta (L+1) - 1 \rfloor)  = \sum_{i = 0}^{\lfloor \delta (L+1) - 1 \rfloor}\binom{L}{i}\delta^i(1-\delta)^{L - i} \nonumber \\
      & = I_{1 - \delta}(L - \lfloor \delta (L+1) - 1 \rfloor, \lfloor \delta (L+1) - 1 \rfloor + 1) \\
      & = I_{1 - \delta}(L + 1 - \lfloor (L + 1)\delta \rfloor, \lfloor (L + 1)\delta \rfloor)
\end{align*}
where $I$ is the incomplete beta function ratio, where $I_{1 - \delta}(L + 1 - \lfloor (L + 1)\delta \rfloor, \lfloor (L + 1)\delta \rfloor)$ is exactly the cumulative distribution function of the Beta distribution with the parameters listed.
\end{proof}
\end{lemma}

As noted already in \cite{lin2024verification}, and as we can see from the proof above, we have in general that
\begin{equation}\label{eq:binom}
    \mathbb{P}_L(\mathbb{P}(R \le C_m) \ge 1 - \delta) \ge 1 - \sum_{i = 0}^{l - 1}\binom{L}{i}\delta^i(1-\delta)^{L - i}
\end{equation}
with  $1 - \sum_{i = 0}^{l - 1}\binom{L}{i}\delta^i(1-\delta)^{L - i} = \sum_{i = l}^L\binom{L}{i}\delta^i(1\!-\!\delta)^{L \!-\! i}$ so that $\mathbb{P}_L(\mathbb{P}(R \le C_m) \ge 1 - \delta) = \sum_{i = l}^L\binom{L}{i}\delta^i(1\!-\!\delta)^{L \!-\! i}$ when $P_R$ is continuous. Equation \eqref{eq:binom} is a conditional guarantee, but we remark that \eqref{eq:binom} is usually a lower bound (e.g. $\delta = 0.1$ and $L = 100$ yield a lower bound confidence of around 0.55).

Based on Theorem \ref{thm:quantile_shift}, we next illustrate the connection between our result and the one-sided Chernoff Bound.\\

\begin{remark}\label{remark:chernoff}
    We first recall the one-sided Chernoff Bound. Given a candidate solution $\hat{x} \in \mathcal{X}$ and a pre-defined confidence level $\beta \in (0, 1)$, it holds that $\mathbb{P}_L(\mathbb{P}(f(\hat{x}, Y) > 0) > \rho) \le \beta,$ where $\rho := \frac{\sum_{i = K + 1}^{K + L}(\mathbbm{1}(f(\hat{x}, Y^{(i)}) > 0)}{L} + \sqrt{\frac{\ln{\beta}}{-2L}}$. Equivalently, we can write this guarantee as
    \begin{align}\label{eq:shift_cond_n}
        \mathbb{P}_L(\mathbb{P}(f(\hat{x}, Y) \le 0) \ge 1 - \delta_n^*) \ge 1 - \beta,
    \end{align}
    where $\delta_n^* \coloneq 1 - \frac{S}{L} + \sqrt{\frac{\ln(1/\beta)}{2L}}$ and $S \coloneq \sum_{i = K + 1}^{K+L}\mathbbm{1}(f(\hat{x}, Y^{(i)}) \\ \le 0)$. Since our result in Theorem \ref{thm:quantile_shift} holds for any feasible solution $\hat{x}$ of the CPP problem \eqref{eq:cp_quantile}, we can compare the guarantee in \eqref{eq:shift_cond_n} from \cite{shang2020posteriori} with our guarantee in \eqref{eq:shift_cond_2} and note that  $\lim_{L\to \infty} \delta_n^* = \lim_{L\to \infty} \delta^*$.\\
\end{remark}

\end{document}